\documentclass[a4paper,11pt]{article}
\pdfoutput=1 

\usepackage{jcappub} 

\usepackage[T1]{fontenc} 

\usepackage{graphicx}
\usepackage{caption}
\usepackage{xcolor}
\usepackage{textcomp}
\usepackage{cancel}
\usepackage[labelformat=parens]{subcaption}
\usepackage{array,multirow}
\usepackage{slashed}
\usepackage{tabularx}
\usepackage{hyperref}
\usepackage{amsmath}
\usepackage{amssymb}
\usepackage{textcomp}
\usepackage{array,multirow}
\usepackage{amssymb}
\usepackage{lineno,hyperref}
\usepackage{lineno}
    \nolinenumbers
\title{\boldmath Triplet scalar flavored leptogenesis with spontaneous CP violation}

\author[a]{Sreerupa Chongdar,}
\author[b,1]{Sasmita Mishra
\note{Corresponding author.}}

\affiliation[a]{Department of Physics and Astronomy, National Institute of Technology Rourkela,\\Sundargarh, Odisha, India, 769008}
\affiliation[b]{Department of Physics and Astronomy, National Institute of Technology Rourkela,\\Sundargarh, Odisha, India, 769008}

\emailAdd{518PH1002@nitrkl.ac.in}
\emailAdd{mishras@nitrkl.ac.in}

\abstract{The inclusion of two triplet scalars in the Standard Model (SM) enables to accommodate neutrino mass generation as well as baryogenesis through leptogenesis. One of the
essential ingredients of leptogenesis is the violation of 
charge conjugation and parity (CP) symmetry in lepton number violating decays of the triplet scalars. We work on the promising sector of spontaneous CP violation (SCPV) which is manifested by the involvement of one scalar singlet and two scalar fields, added to the SM. The predictive
aspect of the model is accomplished by imposing $A_4 \times Z_4$ symmetry which results in the traditional tribimaximal mixing pattern. With updated data on neutrino oscillation, we study the parameter space of the model.
The phase of the complex vacuum expectation value (VEV) of the singlet scalar acts as the common source of
CP violation in both low and high energy sectors.
Due to the flavor symmetry of the model, required baryon asymmetry cannot be accomplished via unflavored leptogenesis. In the temperature regime, $\left[ 10^{9}, 10^{12} \right]$ GeV when flavor effects become important in the study of leptogenesis, it is shown that baryogenesis is achievable. The rich flavor interplay is explored through the study of the density matrix equations. We also study the interplay of hierarchical branching ratios of the decay of the triplet scalars and SCPV phase to accommodate the required CP asymmetry to account for the final baryon asymmetry in the observational range. Considering all possible mass hierarchies among the triplet scalars, the flavor structure of the triplet Yukawa couplings results in different scales of leptogenesis.}
\begin{document}
\maketitle
\flushbottom
\section{Introduction}
The observations from Big Bang nucleosynthesis (BBN) and cosmic microwave
background (CMB) quantify the baryon asymmetry of the Universe (BAU)  
\cite{WMAP:2003elm, WMAP:2008lyn,WMAP:2008ydk, 
Dunkley:2008ie,Planck:2015fie, Planck:2018vyg}, 
$\eta_B=\frac{(n_B-n_{\overline{B}})}{n_\gamma}$ as  
$4.7\times 10^{-10}\leq\eta_{B}\leq 6.5\times 10^{-10}$,
where $n_{B}$, $n_{\overline{B}}$, and $n_{\gamma}=
\frac{2\zeta(3)}{\pi^{2}}T^{3}$ denote the number densities 
of baryons, antibaryons and gamma photons, respectively.
Baryogenesis through leptogenesis \cite{Fukugita:1986hr}
 has become a popular theory to explain the matter-antimatter asymmetry of the Universe. Concurrently, thermal leptogenesis could be an important consequence of neutrino mass generation via the seesaw mechanism, since the heavy beyond Standard Model (BSM) particles introduced through the seesaw mechanism could be promising candidates for generating lepton asymmetry. This is how a seesaw framework could explain tiny non-zero neutrino mass and the consequent leptogenesis leading to adequate baryon asymmetry. Apart from the widely studied type-I seesaw mechanism with heavy right-handed neutrinos, type-II and type-III seesaw mechanism with scalar and fermion triplets, respectively, arise naturally from Weinberg's dimension-5 operator of neutrino mass generation. In this paper, we study baryogenesis
 through flavored leptogenesis from the decay of triplet Higgs scalar with spontaneous CP violation along with light neutrino mass generated 
 via type-II seesaw mechanism.
 
 Neutrino mass generation through type-II seesaw mechanism \cite{Magg:1980ut,Lazarides:1980nt,Mohapatra:1980yp,Cheng:1980qt}  and leptogenesis 
 have widely been studied for their unique features coming from the out-of-equilibrium decay of the heavy triplet scalars \cite{Hambye:2012fh, Ma:1998dx,Guo:2004mp, Chao:2007rm}.  Although the addition of just one triplet scalar to the SM can effectively explain the generation of neutrino mass, for producing non-vanishing CP asymmetry for leptogenesis only one triplet scalar is not enough. While a minimal purely triplet leptogenesis can be implemented through two heavy triplet scalars \cite{Hambye:2005tk, Felipe:2013kk, Sierra:2014tqa}, there are more interesting schemes like type-I+II seesaw leptogenesis with one additional right-handed neutrino along with a triplet scalar \cite{Gu:2006wj}. In the case of triplet leptogenesis, the flavor interplay comes to be very crucial \cite{Lavignac:2015gpa} due to
 the unique decay channels of the scalar triplets. While flavored leptogenesis has already been studied in the framework of explicit CP violation \cite{Chongdar:2021tgm}, it is equally appealing to accommodate flavored leptogenesis through SCPV \cite{Lee:1973iz, Kobsarev:1974hf,Szymacha:1974cx,Branco:2012vs,Pramanick:2022put}. In this work, we explore the role of the SCPV via the complex
 VEV of the scalar sector in flavored leptogenesis.
 
Baryogenesis through leptogenesis from the decay of triplet Higgs has some unique attributes as compared to the other variants of the leptogenesis mechanism. They can be realized from the requirement of CP violation and out-of-equilibrium conditions. One important ingredient of leptogenesis, CP violation, in the case of scalar triplet leptogenesis, can come from two possible sources: (i) Explicit CP violation sourced from the complex Yukawa couplings of the triplet Higgs with leptons and complex scalar coupling to the SM Higgs, (ii) Spontaneous CP violation sourced from the complex VEV of a scalar.
  The latter one is more economical in terms of parameter counting and attractive from a theoretical point of view \cite{Branco:2012vs}. 
 In this case, the CP violation necessary for leptogenesis and potentially observable low-energy leptonic CP violation come from a common source. 
 Also, in models of spontaneously broken CP symmetry, extended by triplet scalars, specific to our case, CP symmetry is imposed at the Lagrangian level,
making the Yukawa couplings real. In this case, spontaneous CP violation can be realized through the complex VEVs developed by the triplet scalars of the model. The light neutrino mass matrix being determined by the complex VEV of the triplets and real Yukawa matrices, CP violation can be established at the low scale (such as neutrino oscillation experiments). The high energy CP violation required for the leptogenesis mechanism cannot be accommodated, in this case. However, the model can be extended by singlet scalars to account for both low and high energy CP violation \cite{Branco:2012vs}. Some recent studies by adding scalar singlets to the SM that can give rise to spontaneous CP violation and light neutrino mass have been performed in Ref. \cite{Barreiros:2020gxu} in the context of dark matter and in Ref. \cite{Barreiros:2022fpi}, in the context of
leptogenesis in vanilla type-I seesaw.
 
The coupling of the triplet scalars to the leptons and the SM Higgs scalar
 leads to two different decay channels of the triplet scalar and the lepton number is violated only if both types
of decays co-exist. The interplay of the strength of the respective 
branching ratios, $B^L$ and $B^\phi$ plays an important role in 
successful baryogenesis through leptogenesis, where $B^L (B^\phi)$ corresponds 
to the branching ratios of the triplet scalar to the leptons (SM Higgs scalar).
The amount of CP asymmetry, $\epsilon$, and the efficiency of leptogenesis depend on $B^L$ and $B^\phi$ in
a different manner. The efficiency is maximal when $B^L \gg B^\phi $ or 
$B^\phi \gg B^L $, whereas this condition corresponds to suppressed CP violation
(as $\epsilon \propto\sqrt{B^L B^\phi}$). Solving a full set of 
Boltzmann equations, in a model-independent way, it was
shown in Ref. \cite{Hambye:2005tk} that even for extremely hierarchical branching ratios, 
it is possible to achieve the required BAU. In this work, we focus on leptogenesis from the decay of triplet scalars with hierarchical branching ratios
and CP violation originating from the VEV of a scalar singlet, embedded in the SM along with two triplet scalars. Also to make the model more predictive we impose $A_4$ symmetry at the Lagrangian level. Since in our case, 
adequate baryon asymmetry is obtained for the triplet scalar mass around $\sim 10^{10}$ GeV, flavor effects become important in this regime. Also, considering
hierarchical branching ratios of triplet scalar decays, washout effects play an important role in generating adequate lepton/baryon asymmetry. With the improved 
computation of the kinetic evolution of flavored lepton asymmetries, we use Density Matrix Equations (DME) for the evaluation of final lepton asymmetries. We also study, with all possible mass hierarchies among the triplet scalars, the flavor structure of the model predicts
different scales of leptogenesis.
 
The paper is organized as follows.
The model is introduced in section (\ref{sect-Model}).
Section (\ref{sec:bau}) is devoted to point out the important features of flavored CP asymmetry and DME for the analysis.
Finally, the results are presented in section (\ref{sec:results}) with the summary. The conclusion is given in section (\ref{sect-conc}).
Three appendices (\ref{App:A}), (\ref{app:rates}), (\ref{app:epsilons}) are added containing some of the detailed calculations.
\begin{table}[ht!]
 \centering 
 \begin{tabular}{ |p{2.0cm}||p{1.2cm}||p{2.15cm}||p{0.8cm}||p{0.8cm}||p{1cm}||p{0.8cm}||p{0.8cm}||p{0.8cm}|  }
 \hline
 Field& $L$ &$e_{R},\mu_{R},\tau_{R}$&$\Delta_{1}$&$\Delta_{2}$&$\phi$&$S$&$\Phi$&$\psi$\\
 \hline
 $SU(2)_{L}$& $2$ &$1$&$3$&$3$&$2$&$1$&$1$&$1$\\
 \hline
 $U(1)_{Y}$ & $-1/2$ &$-1$&$1$&$1$&$1/2$&$0$&$0$&$0$\\
 \hline
 $A_{4}$& $3$ &$1,1'',1'$&$1$&$1$&$1$&$1$&$3$&$3$\\
\hline 
 $Z_{4}$& $i$ &$-i$&$1$&$-1$&$i$&$-1$&$i$&$1$\\
 \hline
\end{tabular}
\caption{Field representations under $SU(2)_{L}$, $U(1)_{Y}, A_{4}$ and $Z_{4}$ symmetries.}
\label{table-fieldsrep}
\end{table}
\section{Model Description}
\label{sect-Model}
In this section, we study an extension of the SM, developed in Ref. \cite{Branco:2012vs} to account for SCPV. With updated data from 
neutrino oscillation and cosmology we explore the parameter space of the model. 
In this model, the SM is extended with one complex scalar singlet $S$ and 
two Higgs triplet scalars $\Delta_{a}(a=1,2)$. To generate 
a realistic lepton mixing pattern, two scalar fields, $\Phi$ and $\psi$ 
are added by imposing $A_4$ flavor symmetry and 
$Z_4$ discrete symmetry. The appropriate charge assignments to all the
scalar and fermion fields are summarized in table (\ref{table-fieldsrep}).
The two triplet Higgs fields can be represented in $SU(2)$ representation as,
\begin{equation}
 \Delta_a=
   \begin{pmatrix}
   \Delta^{0}_a & -\frac{\Delta^+_a}{\sqrt{2}} \\
    -\frac{\Delta^+_a}{\sqrt{2}} & \Delta^{++}_a
  \end{pmatrix},
\end{equation}

The product rule to construct singlets from different products of irreducible representations of $A_4$ in Altarelli- Feruglio basis \cite{Altarelli:2010gt, Altarelli:2005yx} is given as
\begin{equation*}
 (a\otimes b)_{1}=a_{1}b_{1}+a_{2}b_{3}+a_{3}b_{2},
\end{equation*}
\begin{equation*}
 (a\otimes b)_{1'}=a_{3}b_{3}+a_{1}b_{2}+a_{2}b_{1},
\end{equation*}
\begin{equation}
 (a\otimes b)_{1''}=a_{2}b_{2}+a_{1}b_{3}+a_{3}b_{1},
\end{equation}
\begin{equation*}
 (a\otimes b)_{3_{s}}=\frac{1}{3}(2a_{1}b_{1}-a_{2}
 b_{3}-a_{3}b_{2},2a_{3}b_{3}-a_{1}b_{2}-a_{2}b_{1},
 2a_{2}b_{2}-a_{1}b_{3}-a_{3}b_{1}),
\end{equation*}
\begin{equation*}
 (a \otimes b)_{3_{a}}=\frac{1}{2}(a_{2}b_{3}-a_{3}b_{2},
 a_{1}b_{2}-a_{2}b_{1},a_{1}b_{3}-a_{3}b_{1}),
\end{equation*}
\begin{eqnarray}
\nonumber
 ((a\otimes b)_{3_{s}}\otimes c)_{1}&=&\frac{1}{3}(2a_{1}b_{1}c_{1}+2a_{2}b_{2}c_{2}
 +2a_{3}b_{3}c_{3}-a_{1}b_{2}c_{3}-a_{1}b_{3}c_{2}\\
 &-&a_{3}b_{1}c_{2}-a_{3}b_{2}c_{1}-a_{2}b_{1}c_{3}-a_{2}b_{3}c_{1}).
\end{eqnarray}
CP invariance is imposed at the Lagrangian level and hence, all the parameters are perceived to be real. 
The CP asymmetry is broken at high energy by the complex VEV of the scalar singlet. Also $A_4 \times Z_4$
symmetry is broken at a high-scale with a specific vacuum configuration leading to the tribimaximal (TBM) lepton mixing.
Below the cutoff scale, $\Lambda$, the effective Yukawa Lagrangian containing the lowest-order term 
($\mathcal{O}(1/\Lambda)$) can be written as, 
\begin{equation}
 \mathcal{L}=\frac{y^{l}_{e}}{\Lambda}(\overline{L}\Phi)_{1}\phi e_{R}+
 \frac{y^{l}_{\mu}}{\Lambda}(\overline{L}\Phi)_{1'}\phi \mu_{R}+
 \frac{y^{l}_{\tau}}{\Lambda}(\overline{L}\Phi)_{1''}\phi \tau_{R}
 +\frac{y_{2}}{\Lambda}\Delta_{2}(L^{T}L\psi)_{1}+\frac{1}{\Lambda'}\Delta_{1}(L^{T}L)_{1}(y_{1}S
 +y'_{1}S^{*})+{ \rm h.c.}.
\end{equation}
The heavy scalar fields develop VEVs along the required directions as
\begin{equation}
 \langle\Phi\rangle =(r,0,0),\quad \langle\psi\rangle =(s,s,s),\quad \langle S \rangle = v_{S}e^{i\alpha}.
\end{equation}
Now the Lagrangian can be written as
\begin{eqnarray}
\nonumber
 \mathcal{L}&=&\frac{y^{l}_{e}r}{\Lambda}\overline{L}_{e}\phi e_{R}+
 \frac{y^{l}_{\mu}r}{\Lambda}\overline{L}_{\mu}\phi \mu_{R}+
 \frac{y^{l}_{\tau}r}{\Lambda}\overline{L}_{\tau}\phi \tau_{R}+
 \frac{y_{2}s}{3\Lambda}\Delta_{2}(2L^{T}_{e}L_{e}+2L^{T}_{\mu}L_{\mu}+
 2L^{T}_{\tau}L_{\tau}-L^{T}_{e}L_{\mu}\\
 \nonumber
 &-& L^{T}_{e}L_{\tau}-L^{T}_{\mu}L_{e}-L^{T}_{\mu}L_{\tau}-L^{T}_{\tau}L_{e}-L^{T}_{\tau}L_{\mu})\\
 \nonumber
 &+&\frac{v_{S}}{\Lambda'}\Delta_{1}(L^{T}_{e}L_{e}+L^{T}_{\mu}L_{\tau}+
 L^{T}_{\tau}L_{\mu})(y_{1}e^{i\alpha}+y'_{1}e^{-i\alpha})+ {\rm h.c.} \\
 &=& Y^{l}_{\alpha\beta} \overline{L}_{\alpha}\phi e_{R\beta}+Y^{\Delta_{1}}_{\alpha\beta}L^{T}_{\alpha}C\Delta_{1}L_{\beta}+Y^{\Delta_{2}}_{\alpha\beta}L^{T}_{\alpha}C\Delta_{2}L_{\beta}+ {\rm h.c.},
\end{eqnarray}
where the Yukawa matrices are 
\begin{equation}
 Y^{l}=
 \begin{pmatrix}
  y_{e} & 0 & 0\\
  0 & y_{\mu} & 0\\
  0 & 0 & y_{\tau}
 \end{pmatrix},
 \quad
 Y^{\Delta_{1}}=y_{\Delta_{1}}
 \begin{pmatrix}
  1 & 0 & 0\\
  0 & 0 & 1\\
  0 & 1 & 0
 \end{pmatrix},
\quad
 Y^{\Delta_{2}}=\frac{y_{\Delta_{2}}}{3}
 \begin{pmatrix}
  2 & -1 & -1\\
  -1 & 2 & -1\\
  -1 & -1 & 2
 \end{pmatrix},
 \label{eqn-Ymatcs}
 \end{equation}
 and
\begin{equation}
 y_{e,\mu,\tau}=\frac{r}{\Lambda}y^{l}_{e,\mu,\tau},
 \quad
 y_{\Delta_{1}}=\frac{v_{S}}{\Lambda'}(y_{1}e^{i\alpha}+y'_{1}e^{-i\alpha}),
 \quad 
 y_{\Delta_{2}}=\frac{y_{2}}{\Lambda}s.
\end{equation}
It can be seen that the Yukawa coupling matrices $Y^{\Delta_{1}}$ and $Y^{\Delta_{2}}$ 
exhibit $\mu-\tau$ symmetry and magic symmetry, respectively.
We assume the singlet scalar $S$ to be very heavy and to have decoupled at a higher 
energy scale than electroweak (EW) and mass scales of the Higgs triplets.
The complex scalar $S$ plays a vital role in this model since the CP symmetry,
otherwise conserved in the Lagrangian level, is spontaneously broken by the complex VEV 
of the scalar singlet. Hence, it is crucial to analyze the scalar
potential for $S$. The detail of the analysis can be found in Appendix (\ref{App:A}).
The minimization of the potential leads to the following conditions,
\begin{equation}
 v^{2}_{S}=\frac{-2\lambda'_{S}m^{2}_{S}+\lambda''_{S}\mu^{2}_{S}}{4\lambda_{S}\lambda'_{S}-
 8\lambda'_{S}-\lambda''_{S}},\quad \cos(2\alpha)=-\frac{\mu^{2}_{S}
 +\lambda''_{S}v^{2}_{S}}{4\lambda'_{S}v^{2}_{S}}.
\end{equation}
To show that the last solution, which also leads in this case to the spontaneous
breaking of the CP symmetry, corresponds to the global minimum of the potential, 
for a particular choice of parameter space such as
$m^{2}_{S}<0$, $\lambda''_{S}\simeq 0$, $\mu_{S}\simeq 0$ and $\lambda_{S}>2\lambda'_{S}>0$ 
and we obtain from Eq.(\ref{eqn-vs2}),
for case-3
\begin{equation}
 v^{2}_{S}\simeq -\frac{m^{2}_{S}}{2(\lambda_{S}-2\lambda'_{S})},\quad \alpha\simeq \pm\frac{\pi}{4},
\end{equation}
which leads to
\begin{equation}
 V_{0}\simeq -\frac{m^{4}_{S}}{4(\lambda_{S}-2\lambda'_{S})},
\end{equation}
which is the absolute minimum of the potential.

One of the constraints on models with spontaneous CP violation comes from
the cosmological scenario of domain wall problem \cite{Zeldovich:1974uw}. 
A review of topological defects like cosmic strings, domain walls, and monopoles arising from cosmological phase transition in the early universe can be found in Ref. \cite{Vilenkin:1984ib}.
The spontaneous breaking
of a global discrete symmetry leads to the formation of domain walls \cite{Kibble:1976sj, Kibble:1980mv}. The energy density of the domain walls redshifts slower than those of matter and radiation. Long-lived domain walls conflict with standard big bang cosmology. There are several solutions to make them unstable. One of the solutions is assuming gravity breaks global discrete
symmetries explicitly, which lifts the degeneracy of the two vacua, there will be no domain wall problem \cite{Rai:1992xw, Dvali:1995cc}. Lifting the degeneracy of the two vacua by an amount $ \epsilon $ generates a pressure difference between the two 
sides of the walls with a tendency to push the wall into the false vacuum. By making a careful analysis of the dynamics of evolution with domain walls in a radiation-dominated universe, it can be shown that the domain walls are pushed to false vacuum if
\begin{equation}
 \epsilon \ge \frac{v_{\rm dis}^6}{M^2_{\rm Pl}},
 \label{eq:energy-diff}
\end{equation}
where $v_{\rm dis}$ is the scale of discrete symmetry breaking and $M_{\rm Pl} \sim 10^{19}$ GeV. In our case, $v_{\rm dis}$ can be replaced by the SCPV scale, $v_S$. The authors of Ref. \cite{Rai:1992xw} have considered generating such gravity-induced terms and their effects in destabilizing the domain walls. 
For a theory with generic neutral scalar field $\phi$ the Planck-scale suppressed/gravity-induced operators can be written as
\begin{equation}
 \Delta V = \frac{c_5}{M_{\rm Pl}} \phi^5 + \frac{c_6}{M^2_{\rm Pl}} \phi^6 + 
 {\rm Higher~ order~ terms}.
 \label{eq:planck-terms}
\end{equation}
In general, the coefficients $c_n$ are of order $1$. The terms on the right-hand side of Eq. (\ref{eq:planck-terms}) are generic. In a specific model such as the one studied in this paper, the effective higher degree potential induced by gravity has to be constructed carefully keeping gauge invariance intact.
In the model under consideration, the Lagrangian has been written keeping 
$A_4 \times Z_4$, CP and gauge invariance intact. Also, we have assumed that the singlet scalar $S$ is very heavy and decouples from the theory at an energy scale much higher
than the electroweak and Higgs-triplet scales. In Appendix (\ref{App:A}), the minimization of the scalar potential is done by taking only the complex vev of $S$ into account.
Keeping these in mind, the first higher-dimensional operator that can break 
CP symmetry explicitly can be written in terms of the complex gauge singlet  as 
\begin{equation}
 \Delta V = \frac{c_6}{M^2_{\rm Pl}} (SS^*)^3 + {\rm H.c.}+
 {\rm Higher~ order~ terms}.
\end{equation}
Since gravity is expected to break discrete symmetry explicitly, CP symmetry breaking can be 
ensured by taking $c_6$ as complex. However, the successful removal of domain walls requires $c_6 \gtrsim 1$ as per Eq.(\ref{eq:energy-diff}). So careful study of the detailed consequences of gravity is important in low energy issues such as CP violation. Nevertheless, gravity can break the $Z_4$ discrete symmetry ($S$ is charged under $Z_4$). Thus, in this case, it is possible to construct a dimension-five operator of the form
\begin{equation}
 \Delta V = \frac{c_5}{M_{\rm Pl}} \mathcal{O}+ {\rm H.c.}+
 {\rm Higher~ order~ terms},
 \label{eq:dim5}
\end{equation}
where the operator $\mathcal{O}$ can be of the form $S^5, ( S S^*)S^3$, and $(S S^*)^2 S$.
CP symmetry is explicitly broken with the dimension-five operator with $c_5$ as complex. However by comparing the above contribution as per Eq.(\ref{eq:energy-diff}), the relation $c_5 \gg \frac{v_S}{M_{\rm Pl}}$ has to hold. This ensures the dimension-five operator is much bigger than the amount
$\frac{v_S^6}{M^2_{\rm Pl}}$. In our case the scale of leptogenesis is
$> 10^{10}$ GeV. So taking the $v_S$, the SCPV scale above the leptogenesis scale, the values of $c_5$ are bounded below as
\begin{equation}
c_5 > 
 \begin{cases}
  10^{-7}, & v_S \sim 10^{12}, \\
  10^{-6}, & v_S \sim 10^{13},\\
  10^{-5}, & v_S \sim 10^{14}.
 \end{cases}
\end{equation}
The above bound is easily satisfied as $c_5$ is of order $1$.
This makes the domain wall network disappear early enough for an
acceptable leptogenesis scenario. 

There are several other ways to deal with the domain wall problem.
The domain walls can become unstable by the existence of a non-zero $\theta_{\rm QCD}$ through non-perturbative effects,
that break the CP explicitly and lift the degeneracy of the two CP conjugate vacua \cite{Krauss:1992gf}. Another idea in the same direction is to invoke inflation after the domain walls are produced \cite{Langacker:1987ft}. 
In Ref. \cite{McDonald:1997vy}, the authors study the possibility of spontaneous discrete symmetry breaking 
due to a gauge singlet field whose initial value is fixed during inflation, so that the resulting discrete symmetry-breaking phase is the same over the whole observable universe.
However, the baryon asymmetry has to be generated after inflation to avoid diluting it to an insignificant level. In most of the models on leptogenesis, the reheat temperature required is $10^{12}$ GeV (in our case the scale of leptogenesis is $\mathcal{O}(10^{10})$ GeV).
So it is hard to make domain walls unstable unless they are associated 
with a spontaneous breaking $> 10^{12}$ GeV which fits very well in our case. 

The various ways to solve domain wall problem show that it is possible to overcome it independently.
In all the scenarios considered above, it does not prevent the complex VEV of $S$
in generating a complex Yukawa coupling required for CP violation in leptogenesis.

The triplet Higgs scalars contribute to the neutrino mass generation 
through the type-II seesaw mechanism as
\begin{equation}
 m_{\nu}=m^{(1)}_{\nu}+m^{(2)}_{\nu},
 \quad m^{(a)}_{\nu}=2u_{a}Y^{\Delta_{a}},
 \label{eqn-neumass}
\end{equation}
where $m^{(1)}_{\nu}$ and $m^{(2)}_{\nu}$ are the contributions coming from 
the triplet scalars $\Delta_{1}$ and $\Delta_{2}$, respectively. Here, $u_{a}$ are 
the VEVs of the neutral components of the triplet scalars $\Delta_{a}$, and it is given by
\begin{equation}
 u_{a}=\mu^{*}_{a}v^{2}/M_{a}, \quad 
 \mu_{1}=(\lambda_{1}e^{i\alpha}+\lambda'_{1}e^{-i\alpha})v_{S}/M_{1}
 \label{eqn-muexp}
\end{equation}
where $v= \langle \phi_{0} \rangle =174$ GeV.
To diagonalize the neutrino mass matrix, we define a unitary matrix $U$ as
\begin{equation}
 m_{\nu}=U^{*}d_{\nu}U^{\dagger},\quad 
 d_{\nu}=\rm{diag}(|z_{1}e^{i\beta}+z_{2}|,z_{1},|z_{1}e^{i\beta}-z_{2}|)\equiv
 diag(m_{1},m_{2},m_{3}),
 \label{eqn-mnudiag}
\end{equation}
where
\begin{equation}
 z_{a}=2|u_{a}y_{\Delta_{a}}|,
 \quad \beta={\rm arg}(u_{1}y_{\Delta_{1}})={\rm arctan}\left(\frac{\lambda'_{1}-
 \lambda_{1}}{\lambda'_{1}+\lambda_{1}}{\rm tan}\alpha\right)+{\rm arctan}
 \left(\frac{y'_{1}-y_{1}}{y'_{1}+y_{1}}{\rm tan}\alpha\right).
 \label{eqn-zabeta}
\end{equation} 
In this model, where the neutrino mass matrix can be decomposed with 
matrices with $\mu-\tau$ symmetry and magic symmetry, the mixing matrix 
takes the form
\begin{equation}
 U=e^{-i\sigma_{1}/2}U_{\rm{TBM}}K,
\end{equation}
where $U_{\rm{TBM}}$ is the tribimaximal mixing matrix and is given by,
\begin{equation}
 U_{\rm{TBM}}=
 \begin{pmatrix}
  \frac{2}{\sqrt{6}} & \frac{1}{\sqrt{3}} & 0\\
  -\frac{1}{\sqrt{6}} & \frac{1}{\sqrt{3}} & -\frac{1}{\sqrt{2}}\\
  -\frac{1}{\sqrt{6}} & \frac{1}{\sqrt{3}} & \frac{1}{\sqrt{2}}
 \end{pmatrix},
\end{equation}
and the Majorana phases are the entries in the diagonal matrix $K$, given by,
\begin{equation}
 K=\rm{diag}(1,e^{i\gamma_{1}},e^{i\gamma_{2}}),
\end{equation}
with 
\begin{equation}
 \gamma_{1}=(\sigma_{1}-\beta)/2,
 \quad \gamma_{2}=(\sigma_{1}-\sigma_{2})/2,
 \quad \sigma_{1,2}=\rm{arg}(z_{2}\pm z_{1}e^{i\beta}).
 \label{eqn-gm12}
\end{equation}
TBM mixing matrix has gathered considerable attention because of its proximity with the observation of 
neutrino masses and mixing. However, the absence of Dirac-type CP violating phase asks for possible modifications 
of the model, which we discuss in the next subsection.
Considering the two cases of neutrino mass hierarchy, we can have two sets 
of constraints on $z_{1}$ and $z_{2}$. From Eq.(\ref{eqn-mnudiag}), 
for normal hierarchy (NH), i.e. $m_{3}>m_{2}>m_{1}$, we obtain
\begin{equation}
 z_{2}+2z_{1}\cos\beta<0, 
 \quad
 z_{2}-2z_{1}\cos\beta>0,
 \label{eqn-zNH}
\end{equation}
and, for inverted hierarchy (IH) we have 
\begin{equation}
 z_{2}+2z_{1}\cos\beta<0, 
 \quad 
 z_{2}-2z_{1}\cos\beta<0.
 \label{eqn-zIH}
\end{equation}
Now the constraints in Eq.(\ref{eqn-zIH}) 
cannot be satisfied simultaneously, which indicates that the model under 
consideration does not qualify for the IH of neutrino mass. 
In terms of the phase $\beta$ and two mass squared differences, 
the parameters $z_{1}$ and $z_{2}$ can be written as,
\begin{equation}
z_{1}=-\frac{1}{2\cos\beta}\frac{\Delta m^{2}_{31}}{\sqrt{2(\Delta m^{2}_{31}-
2\Delta m^{2}_{21})}}\simeq -\frac{1}{2\cos\beta}\sqrt{\frac{\Delta m^{2}_{31}}{2}},
\label{eqn-calz1}
\end{equation}
and
\begin{equation}
 z_{2}=\sqrt{\frac{\Delta m^{2}_{31}-2\Delta m^{2}_{21}}{2}}\simeq \sqrt{\frac{\Delta m^{2}_{31}}{2}}.
 \label{eqn-calz2}
\end{equation}
Hence, the neutrino mass eigenvalues can be expressed as 
\begin{equation}
 m_{1}=\sqrt{z^{2}_{1}-\Delta m^{2}_{21}}, 
 \quad m_{2}=z_{1},
 \quad m_{3}=\sqrt{z^{2}_{1}+\Delta m^{2}_{31}-\Delta m^{2}_{21}}.
 \label{eqn-calm123}
\end{equation}
\subsection{Non-zero $\theta_{13}$ from perturbative corrections in the flavon VEV}
\label{sect-TBM} 
The measurement of non-zero $\theta_{13}$ \cite{T2K:2011ypd,MINOS:2011amj,DayaBay:2012fng} requires to modify the TBM scenario.
It can be achieved by exploiting the small 
perturbations around the TBM vacuum alignment conditions. More specifically,
by considering a small perturbation around the flavon VEV of the form $\langle \Phi \rangle = r(1,\varepsilon_{1},\varepsilon_{2})$ with $|\varepsilon_{1,2}|\ll1$
we obtain the transformed charged lepton Yukawa matrix,  
\begin{equation}
 Y^{l}=
 \begin{pmatrix}
  y_{e} & y_{\mu}\varepsilon_{1} & y_{\tau}\varepsilon_{2}\\
  y_{e}\varepsilon_{2} & y_{\mu} & y_{\tau}\varepsilon_{1}\\
  y_{e}\varepsilon_{1} & y_{\mu}\varepsilon_{2} & y_{\tau}
 \end{pmatrix},
 \label{eqn-Yl}
\end{equation}
while the lepton mixing coming from the neutrino sector remains invariably of 
TBM type, since $\langle \psi \rangle$ is not perturbed.
The emergence of a non-diagonal charged
lepton Yukawa matrix can be diagonalized by biunitary transformation
$ Y^{l}= U_{l} ~{\rm Diagonal} (y_e, y_\mu, y_\tau) ~I$, where $I$ is the identity matrix and $U_{l}$ is given by
\begin{equation}
 U_{l}=
 \begin{pmatrix}
  1 & \varepsilon_{1} & \varepsilon_{2}\\
  \varepsilon_{2} & 1 & \varepsilon_{1}\\
   \varepsilon_{1} & \varepsilon_{2} & 1
  \end{pmatrix}. 
  \label{eqn-Uprt}
\end{equation}
Further, the lepton mixing matrix gets modified as 
\begin{equation}
 U=e^{-i\sigma_{1}/2}U^{\dagger}_{l}U_{TBM}K.
 \label{eqn-pTBM}
\end{equation}
The $U$ matrix obtained in Eq. (\ref{eqn-pTBM}) can be identified as Pontecorvo-Maki-Nakagawa-Sakata (PMNS) mixing matrix, $U_{{\rm PMNS}}$.
\subsection{Parameter space study}
\label{sect-Paraspace}
To check the compatibility of the model with the latest experimental results, it is
important to study the parameter space of the model constrained by the
neutrino oscillation data. The latest global-fit of the observational data
are summarized in table (\ref{table-neop22}) \cite{Esteban:2020cvm}. 
Also, the limits for effective mass, $|m_{ee}|$ appearing in the half-life of 
neutrinoless double beta ($0\nu\beta\beta$) decay are constrained by GERDA as $<(79-180)$ meV \cite{GERDA:2020xhi} and KamLAND-Zen as $<(36-156)$ meV \cite{KamLAND-Zen:2022tow}. The future sensitivity of $|m_{ee}|$ from the future ton-scale experiments like CUPID \cite{CUPID:2019imh,Armengaud:2019loe}, LEGEND \cite{LEGEND:2017cdu}, and nEXO \cite{nEXO:2017nam,nEXO:2018ylp} are going to probe  $|m_{ee}| \le 0.01$ eV. 
 \begin{table}[t]
 \centering 
 \begin{tabular}{ |p{5.5cm}||p{4.0cm}||p{4.0cm}|  }
 \hline
 \multicolumn{3}{|c|}{NuFIT 5.0 \cite{Esteban:2020cvm}} \\
 \hline
 Neutrino Oscillation parameters& Normal ordering 1$\sigma$ &Normal ordering 3$\sigma$\\
 \hline
 \hline
 $\sin^{2}\theta_{12}$   & $0.303^{+0.012}_{-0.012}$    &$0.270\rightarrow0.341$\\
 $\sin^{2}\theta_{23}$&   $0.451^{+0.019}_{-0.016}$  & $0.408\rightarrow0.603$   \\
 $\sin^{2}\theta_{13}$ &$0.02225^{+0.00056}_{-0.00059}$ & $0.02052 \rightarrow 0.02398$\\
 $\delta/^{\circ}$    &$232^{+36}_{-26}$ & $144\rightarrow350$\\
 $\frac{\Delta m^{2}_{21}}{10^{-5}{\rm eV}^{2}}$&   $7.41^{+0.21}_{-0.20}$  & $6.82\rightarrow8.03$\\
 $\frac{\Delta m^{2}_{3l}}{10^{-3}{\rm eV}^{2}}$& $+2.507^{+0.026}_{-0.027}$  & $+2.427\rightarrow+2.590$   \\
 \hline
\end{tabular}
\caption{Allowed ranges of neutrino oscillation parameters for normal mass ordering.}
\label{table-neop22}
\end{table}
The three mixing angles,
$\theta_{12}$, $\theta_{23}$ and $\theta_{13}$ can be extracted from the standard parameterization of the $U_{\rm PMNS}$ matrix as 
\begin{equation}
 \sin^{2}\theta_{12}=\frac{|U_{12}|^{2}}{1-|U_{13}|^{2}}, \quad
 \sin^{2}\theta_{23}=\frac{|U_{23}|^{2}}{1-|U_{13}|^{2}}, \quad
 \sin^{2}\theta_{13}=|U_{13}|^{2}.
 \label{eqn-sinU}
\end{equation}
From equations (\ref{eqn-pTBM}) and (\ref{eqn-sinU}), three mixing angles,
$\theta_{12}$, $\theta_{23}$ and $\theta_{13}$ are related to the 
small perturbations $\varepsilon_{1}$ and $\varepsilon_{2}$ through the $U$ 
matrix, given in Eq. (\ref{eqn-pTBM}).
The explicit relations relating the three mixing angles
and the small perturbations are shown below: 
\begin{equation}
 \sin^{2}\theta_{12}\simeq\frac{1}{3}[1+2(\varepsilon_{1}+\varepsilon_{2})],\quad
 \sin^{2}\theta_{23}\simeq\frac{1}{2}[1-2\varepsilon_{2}],\quad
 \sin^{2}\theta_{13}\simeq\frac{1}{2}(\varepsilon_{1}-\varepsilon_{2})^{2}, 
\end{equation}
which further satisfies the relation,
\begin{equation}
 \sin^{2}\theta_{13}\simeq\frac{(4\sin^{2}\theta_{23}-3\cos^{2}\theta_{12})^{2}}{8}.
\end{equation}
\begin{figure} [t]
 \centering
   \begin{subfigure}{0.45\linewidth}
   \includegraphics[width=\linewidth]{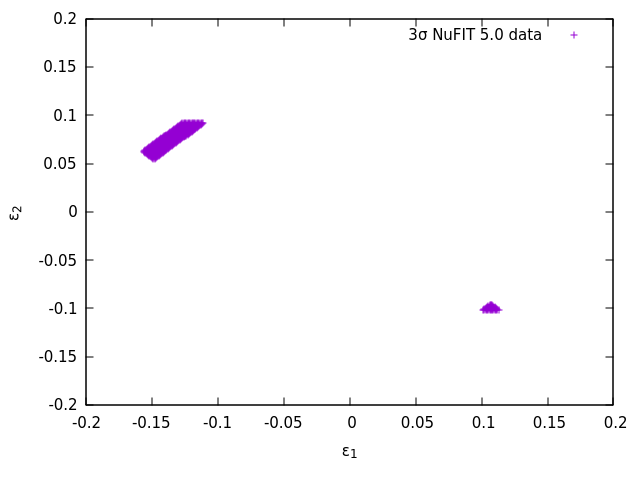}
   \caption{}
   \label{fig-caseAmod}
   \end{subfigure}
   \begin{subfigure}{0.45\linewidth}
   \includegraphics[width=\linewidth]{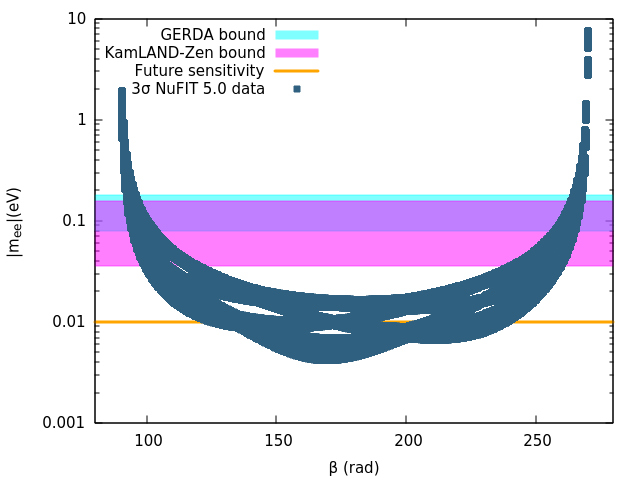} 
   \caption{}
   \label{fig-meeAmod1}
   \end{subfigure}
   \caption{Left: The plot shows the parameter space for the perturbations $\varepsilon_{1}$ 
   and $\varepsilon_{2}$ constrained by the $3\sigma$ data of three mixing 
   angles. Right: The plot shows the dependence of $|m_{ee}|$ (eV) on the phase $\beta$ (rad) 
   using the bounds of the $3\sigma$ data of three mixing angles and the two mass squared differences. The experimental limits on $|m_{ee}|$ are given by GERDA (cyan) and KamLAND-Zen (magenta). The orange line corresponds to the future sensitivity of $|m_{ee}|$ from the future ton-scale experiments like CUPID, LEGEND, and nEXO. }
\end{figure}
By imposing the experimentally obtained constraints on the mixing angle, as 
given in table (\ref{table-neop22}), the allowed parameter 
space for the perturbations $[\varepsilon_{1},\varepsilon_{2}]$ are obtained 
as shown in Fig. (\ref{fig-caseAmod}). We use the same set of 
mixing angle constraints along with the bounds on two mass squared differences to assess the effective neutrino mass, which is given by
\begin{equation}
 |m_{ee}|=\frac{1}{3}\left|2m_{1}(1-\varepsilon_{1}-\varepsilon_{2})
 +m_{2}(1+2\varepsilon_{1}+2\varepsilon_{2})e^{2i\gamma_{1}}\right|,
 \label{eq:mee}
\end{equation}
where the phase $\gamma_{1}$ is given in Eq. (\ref{eqn-gm12}).
Through the phase, $\gamma_{1}$, it is possible to draw a link between the
effective mass  and its dependence on the phase $\beta$, 
which appears in the expressions of neutrino masses as well as in the high-scale CP violation, as it will be elaborated later on. In Fig. (\ref{fig-meeAmod1}) 
we show the dependence of $|m_{ee}|$ on the phase $\beta$. 
In generating the plot using Eq.(\ref{eq:mee}), we have calculated the masses,
$m_1, m_2$ and $m_3$ using Eq.(\ref{eqn-calm123}), which requires the use of equations (\ref{eqn-calz1}) and (\ref{eqn-calz2}). Also, we ensure the values of $z_a$ to be  positive as per Eq.(\ref{eqn-zabeta}).
The plot shows that the allowed effective mass in turn has imposed a bound 
on the phase $\beta$ $\sim (100^{\circ} - 260^{\circ})$. We have used this restricted parameter space of the phase $\beta$ to calculate the flavored CP asymmetries in section (\ref{sec:results}).
\section{Baryogenesis through leptogenesis}
\label{sec:bau}
The Sakharov conditions \cite{Sakharov:1967dj} provide necessary
ingredients for    dynamical generation of BAU viz; baryon number
 violation, C and CP violation and out-of-equilibrium dynamics in the processes involving baryons.
In the framework of baryogenesis through leptogenesis, the necessary CP violation required for successful 
leptogenesis is provided by the decay of heavy particles added to the SM. In the type-II seesaw scenario, the decays of the triplet scalars 
$\Delta_{1}$ and $\Delta_{2}$ serve the purpose of generating adequate CP 
asymmetry. 

The flavor-independent CP asymmetry, as shown in Eq. (\ref{eqn-cpunf}),
crucially depends on the factor ${\rm Tr}(m^{(a)}_{\nu}m^{(b)\dagger}_{\nu})$ that comes
from the neutrino mass matrices given in Eq. (\ref{eqn-neumass}). In the framework under
consideration, this factor vanishes due to the specific flavor structure of the 
Yukawa matrices $Y^{\Delta_{a}}$, causing the flavor-independent CP asymmetry 
to diminish. Hence, the flavor-independent analysis of leptogenesis is not 
feasible in our chosen framework. Moreover, since scalar triplets generally 
couple to lepton pairs from different flavor generations, it is more appropriate
to consider the flavor effects while studying triplet scalar leptogenesis. 

\subsection{CP violation}
\label{sect-LCPV}
\begin{figure}[h]
 \centering
\includegraphics[scale=0.4]{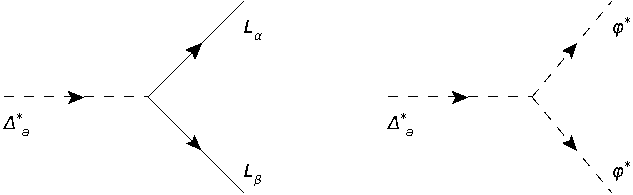}
\caption{Tree-level decay diagram: Triplet scalar decays into lepton and SM Higgs, respectively.}
\label{fig-treelevel}
\end{figure}

\begin{figure}[h]
 \centering
\includegraphics[scale=0.3]{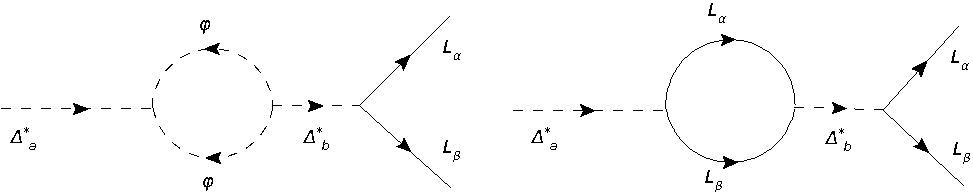}
\caption{One-loop decay diagram: Triplet scalar decays into leptons through Higgs loop and lepton loop, respectively.}
\label{fig-oneloop}
\end{figure}

A triplet scalar can decay into two leptons or two SM Higgs doublets
at tree level. Lepton asymmetry can be produced only if 
the two decay channels co-exist. The non-vanishing 
lepton asymmetry can be produced for each triplet component from  the decay of  
triplet scalar $\Delta_{a}$ into two leptons and the CP asymmetry can be estimated through,
 \begin{equation}
 \epsilon_{a}=\Delta L\sum_{\alpha\beta}\frac{\Gamma(\Delta^{*}_{a}\rightarrow 
 L_{\alpha}+L_{\beta})-\Gamma(\Delta_{a}\rightarrow \overline{L}_{\alpha}+
 \overline{L}_{\beta})}{\Gamma_{\Delta_{a}}+\Gamma_{\Delta^{*}_{a}}},
\end{equation}
where $\Delta L=2$ since the lepton number is violated by two units. 
The tree level branching ratios $B^{L}_{a}$ and $B^{\phi}_{a}$,
\begin{equation}
 B^{L}_{a}\Gamma_{\Delta_{a}}\equiv\sum_{\alpha,\beta}\Gamma(\Delta^{*}_{a}\rightarrow 
 L_{\alpha}+L_{\beta})=\frac{M_{a}}{8\pi}{\rm Tr}(Y^{\Delta_{a}\dagger}Y^{\Delta_{a}}),
 \label{eq:BR-L}
\end{equation} 
\begin{equation}
 B^{\phi}_{a}\Gamma_{\Delta_{a}}\equiv\Gamma(\Delta^{*}_{a}\rightarrow \phi+\phi)=
 \frac{M_{a}}{8\pi}|\mu_{a}|^{2},
 \label{eqn-Blhfunc}
\end{equation}
satisfy the relation $B^{L}_{a}+B^{\phi}_{a}=1$, provided
\begin{equation}
 \Gamma_{\Delta_{a}}=\frac{M_{a}}{8\pi}\left[{\rm Tr}(Y^{\Delta_{a}\dagger}Y^{\Delta_{a}})
 +|\mu_{a}|^{2}\right],
 \label{eqn-gmD}
\end{equation}
which is the total triplet decay width. In the case of a lepton flavor violating 
interaction, the interference of tree level (Fig. \ref{fig-treelevel}) and
one-loop (Fig. \ref{fig-oneloop}) diagrams give rise to a non-vanishing CP asymmetry, 
given by
\begin{equation}
 \epsilon^{\alpha\beta}_{a}\simeq-\frac{g(x_{b})}{2\pi}\frac{c_{\alpha\beta}
 {\rm Im}[\mu^{*}_{a}\mu_{b}Y^{\Delta_{a}}_{\alpha\beta}Y^{\Delta_{b}*}_{\alpha\beta}]}{{\rm Tr}(Y^{\Delta_{a}\dagger}Y^{\Delta_{a}})+|\mu_{a}|^{2}},
 \label{eqn-vab}
\end{equation}
where $x_{b}=\frac{M^{2}_{b}}{M^{2}_{a}}$, and the on-loop self energy function,
\begin{equation}
 g(x_{b})=\frac{\sqrt{x_{b}}(1-x_{b})}{(x_{b}-1)^{2}+(\Gamma_{\Delta_{b}}/M_{a})^{2}},
\end{equation}
 and 
\begin{eqnarray*}
 c_{\alpha\beta}&=&2-\delta_{\alpha\beta}\quad \rm{for}\quad\Delta^{0}_{a},\Delta^{++}_{a},\\
 &=& 1 \quad \rm{for}\quad\Delta^{+}_{a}.
\end{eqnarray*}
In the present scenario, we can rewrite Eq. (\ref{eqn-vab}) as 
\begin{equation}
\epsilon^{\alpha\beta}_{a}\simeq-\frac{g(x_{b})}{4\pi}\frac{M_{b}(B^{L}_{a}B^{\phi}_{a})^{\frac{1}{2}}}{v^{2}}
\frac{c_{\alpha\beta}{\rm Im}[m^{(a)}_{\nu,\alpha\beta}m^{*}_{\nu,\alpha\beta}]}
{[{\rm Tr}(m^{(a)\dagger}_{\nu}m^{(a)}_{\nu})]^{1/2}}. 
\end{equation}
If we consider the triplet scalar masses to be hierarchical, satisfying $M_{a}\ll M_{b}$ then 
\begin{equation}
 \epsilon^{\alpha\beta}_{a}\simeq\frac{M_{a}(B^{L}_{a}B^{\phi}_{a})^{\frac{1}{2}}}{v^{2}}
 \frac{c_{\alpha\beta}{\rm Im}[m^{(a)}_{\nu,\alpha\beta}m^{*}_{\nu,\alpha\beta}]}
 {[{\rm Tr}(m^{(a)\dagger}_{\nu}m^{(a)}_{\nu})]^{1/2}}.
 \label{eqn-vflav}
\end{equation}
We obtain the flavor independent or unflavored CP asymmetry from Eq. (\ref{eqn-vflav}),
by summing over all lepton flavors,
\begin{equation}
 \epsilon_{a}\simeq\frac{M_{a}(B^{L}_{a}B^{\phi}_{a})^{\frac{1}{2}}}{v^{2}}
 \frac{{\rm Im}[{\rm Tr}(m^{(a)}_{\nu}m^{\dagger}_{\nu})]}{[{\rm Tr}(m^{(a)\dagger}_{\nu}m^{(a)}_{\nu})]^{1/2}}.
 \label{eqn-cpunf}
\end{equation}
This CP asymmetry has an upper bound as can be seen from the expression,
\begin{equation}
 |\epsilon_{a}|\leq\frac{M_{a}(B^{L}_{a}B^{\phi}_{a})^{\frac{1}{2}}}{v^{2}}[{\rm Tr}(m^{(a)}_{\nu}m^{\dagger}_{\nu})]^{1/2}=\frac{M_{a}(B^{L}_{a}B^{\phi}_{a})^{\frac{1}{2}}}{v^{2}}\left(\sum_{k}m^{2}_{k}\right)^{1/2}.
\end{equation}
Further, for hierarchical light neutrinos, one finds
\begin{equation}
 |\epsilon_{a}|\lesssim 10^{-6}(B^{L}_{a}B^{\phi}_{a})^{\frac{1}{2}}\left(\frac{M_{a}}{10^{10} GeV}\right)\left(\frac{\sqrt{\Delta m^{2}_{31}}}{0.05 eV}\right)
\end{equation}
which reaches its absolute maxima at $B^{L}_{a}=B^{\phi}_{a}=\frac{1}{2}$.
\subsection{Triplet leptogenesis and flavor covariance}
\label{sect-triplept}
Flavor effects can emerge  from the misalignment of flavor and mass 
eigenbasis, which can persist in the off-diagonal entries of charged lepton
Yukawa matrices. The phenomenon of particle mixing with primordial plasma 
as a background medium also contributes to flavor effects in the early universe.
Such effects can be addressed through quantum statistical mechanics, precisely 
considering the non-vanishing off-diagonal entries of particle number densities 
of particle species tagged with flavor quantum numbers. In the case of thermal
leptogenesis, the charged lepton flavor coherence affects the washout process,
thereby affecting the generation of baryon asymmetry. Hence it is necessary to
develop theoretical frameworks to assess the flavor effects. We will use a 
formalism, namely closed time path (CTP) formalism, in which particle densities
are replaced by Green's functions to describe non-equilibrium phenomena like
thermal leptogenesis \cite{Lavignac:2015gpa, Dev:2017trv, Stodolsky:1986dx, Raffelt:1991ck, Sigl:1993ctk}.
The flavor covariance could be conveniently studied in a density matrix formalism \cite{Lavignac:2015gpa} through flavor-covariant Boltzmann equations.  
The temperature range $[10^{9},10^{12}]$ GeV appears more interesting for the study of such flavor covariance since this is an intermediate temperature range $\lesssim 10^{12}$ GeV, when the flavor coherence between $\tau$-leptons and the other two leptons $(e,\mu)$ is broken, yet before the complete disintegration of the flavor coherence that happens further below $10^{9}$ GeV.

Hence, we will study the leptogenesis through a set of
flavor-covariant Boltzmann equations and perceive the asymmetries in terms of
a density matrix $(\Delta_{l})_{\alpha\beta}$. In general, it is a $3\times3$
matrix in lepton flavor space. The lepton doublet asymmetries are stored as 
diagonal entries, whereas, the off-diagonal entries symbolise the quantum 
correlation between the different flavors. In the most general way, we can 
write such a flavor-covariant equation to express the evolution of flavored lepton asymmetries, as 
\begin{equation}
 sHz\frac{d(\Delta_{l})_{\alpha\beta}}{dz}=\rm{Source}+\rm{Washout}.
\end{equation}
The source term on the right-hand side is proportional to the flavor-covariant
CP asymmetry matrix $\epsilon^{\alpha\beta}$. Different lepton number violating 
processes like inverse triplet and anti-triplet decays washout the lepton 
asymmetry generated through decay processes. Similarly, $\Delta L=2$ 
scatterings via s - and t - channel triplet exchange, 2 - lepton - 2 - lepton 
scatterings, as well as lepton flavor violating 
processes also contribute to affect the total lepton asymmetry. So, they
come under the class of washout processes. To avoid complexity, we will
only stick to the washout associated with inverse decays. Apart from the 
inverse decay and scattering processes, there are different SM reactions, 
such as strong and EW sphaleron process and Yukawa couplings that 
indirectly affect the final asymmetry. They are considered spectator processes.
For a specific temperature range, these spectator processes in thermal 
equilibrium can impose some relations among chemical potentials and thereby
among the asymmetries. As a consequence, different asymmetries appearing in
the Boltzmann equation can be expressed in terms of asymmetries conserved by
all active SM processes.           

We write the evolution of the asymmetries of the particle species under 
consideration in a set of flavor-covariant Boltzmann equations, with respect 
to a factor $z=\frac{M_{a}}{T}$, $T$ is the  temperature of the
Universe, $H(z)$ is the Hubble rate of expansion of the Universe, 
$H(z)=\frac{H_{0}(M_{a})}{z^{2}}$, $H_{0}(T)=\sqrt{\frac{4\pi^{3}}{45}g_{*}}
\frac{T^{2}}{M_{P}}$, $M_{P}=1.22\times 10^{19}$ GeV is the Planck mass, 
$s= \left( \frac{2\pi^{2}}{45} \right) g_{*}T^{3}$ is the total entropy density and
$g_{*}=106.75$. The asymmetries are defined as $\Delta_{x}=
\frac{n_{x}-n_{\overline{x}}}{s}$ where $n_x(n_{\bar{x}})$ is the number
density of the species, $x(\bar{x})$, and $\Sigma_{\Delta}=
\frac{n_{\Delta}+n_{\overline{\Delta}}}{s}$ is the total triplet
number density. Also, $Y_x=\frac{n_{x}}{s}$ is the comoving number density, 
\begin{equation*}
 Y^{\rm eq}_{\Delta}=\frac{45g_{\Delta}}{4\pi^{4}g_{*}}z^{2}K_{2}(z), \quad
 Y^{\rm eq}_{L}=\frac{3}{4}\frac{45\varsigma(3)}{2\pi^{4}g_{*}}g_{L}, \quad
 Y^{\rm eq}_{\phi}=\frac{45\varsigma(3)}{2\pi^{4}g_{*}}g_{\phi},
\end{equation*}
are the equilibrium values, with $g_{\Delta}=1$ for each triplet
component, $g_{L}=2$ and $g_{\phi}=2$, $\varsigma(3)\simeq1.202$, $K_2(z)$ is 
the modified Bessel function. Different reaction densities are elaborated in 
Appendix (\ref{app:rates}).

We are interested to study the evolution of the asymmetries $\Delta_{\phi}$, $\Delta_{\Delta}$ and $(\Delta_{l})_{\alpha\beta}$ through a set of flavor-covariant Boltzmann equations. Within the temperature range $T\subset[10^{9},10^{12}]$ GeV, the $\tau$-Yukawa couplings enter equilibrium and consequently, the $(e,\tau)$, $(\mu,\tau)$,
$(\tau,e)$, $(\mu,e)$ entries of the $3\times 3$ density matrix $(\Delta_{l})_{\alpha\beta}$ come to be zero. So it is convenient 
to assign $2\times 2$ matrix $(\Delta^{0}_{l})_{\alpha\beta}$ $(\alpha,\beta=e,\mu)$ and $\Delta_{l_{\tau}}$ instead of the $3\times 3$ density matrix $(\Delta_{l})_{\alpha\beta}$ with $(\alpha,\beta=e,\mu,\tau)$
to express the asymmetries and quantum correlations 
stored in $e-\mu$ flavored lepton subspace  and the asymmetries stored in $\tau$-lepton, respectively. 
Among all the asymmetries, $\Delta_{\phi}$, $\Delta_{\Delta}$, $\Delta_{l_{\tau}}$, $(\Delta^{0}_{l})_{\alpha\beta}$, only $\Delta_{\Delta}$ is conserved by all the SM interactions. 
Hence, it is necessary to define conserved asymmetries $\Delta_{\tau}$ and $\Delta^{0}_{\alpha\beta}$ (instead of $\Delta_{l_{\tau}}$ and $(\Delta^{0}_{l})_{\alpha\beta}$) along with $\Delta_{\Delta}$ to be used in the Boltzmann equations. The Higgs and flavored lepton asymmetries can be expressed as functions of these conserved asymmetries ($\Delta_{\Delta}$, $\Delta_{\tau}$, $\Delta^{0}_{\alpha\beta}$) using suitable chemical potential relations (discussed in equations (\ref{eqn-chemp1} - \ref{eqn-chemp7})). For the temperature range $T\subset[10^{9},10^{12}]$ GeV, we obtain the relations as \cite{Lavignac:2015gpa} 
\begin{equation}
 (\Delta^{0}_{l})_{\alpha\beta}=\left(\frac{86}{589}{\rm Tr}(\Delta^{0}_{\alpha\beta})
 +\frac{60}{589}\Delta_{\tau}+\frac{8}{589}\Delta_{\Delta} \right)\delta_{\alpha\beta}-\Delta^{0}_{\alpha\beta},
 \label{eqn-Dl1}
\end{equation}
\begin{equation}
 \Delta_{l_{\tau}}=\frac{30}{589} {\rm Tr}(\Delta^{0}_{\alpha\beta})-\frac{390}{589}\Delta_{\tau}
 -\frac{52}{589}\Delta_{\Delta},
 \label{eqn-Dl2}
\end{equation}
and 
\begin{equation}
 \Delta_{\phi}=-\frac{164}{589}{\rm Tr}\left(\Delta^{0}_{\alpha\beta}\right)
 -\frac{224}{589}\Delta_{\tau}-\frac{344}{589}\Delta_{\Delta},
 \label{eqn-Dphi}
\end{equation}
where $\alpha$, $\beta$ describe any two orthogonal in the $(l_{e},l_{\mu})$ flavor subspace now.
Therefore, within the temperature range $T\subset[10^{9},10^{12}]$ GeV, the flavor-covariant
Boltzmann equations are given by
\begin{equation}
 sHz\frac{d\Sigma_{\Delta}}{dz}=-\left(\frac{\Sigma_{\Delta}}{\Sigma^{eq}_{\Delta}}-1\right)\gamma_{D}-2\left(\left(\frac{\Sigma_{\Delta}}{\Sigma^{eq}_{\Delta}}\right)^{2}-1\right)\gamma_{A},
 \label{eqn-DME1}
\end{equation}
\begin{equation}
 sHz\frac{d\Delta^{0}_{\alpha\beta}}{dz}=-\left(\frac{\Sigma_{\Delta}}{\Sigma^{eq}_{\Delta}}-1\right)\gamma_{D}\epsilon^{\alpha\beta}+\tilde{\mathcal{W}}^{D}_{\alpha\beta},
 \label{eqn-DME2}
\end{equation}
\begin{equation}
 sHz\frac{d\Delta_{\tau}}{dz}=-\left(\frac{\Sigma_{\Delta}}{\Sigma^{eq}_{\Delta}}-1\right)\gamma_{D}\epsilon^{\tau\tau}+\tilde{\mathcal{W}}^{D}_{\tau},
 \label{eqn-DME2t}
\end{equation}
\begin{equation}
 sHz\frac{d\Delta_{\Delta}}{dz}=-\frac{1}{2}\tilde{\mathcal{W}}^{D}_{\Delta},
 \label{eqn-DME3}
\end{equation}
where, $\tilde{\mathcal{W}}^{D}$ represent different washout terms associated
with inverse decay processes. In order to avoid complexity of presentation, we 
represent $Y^{\Delta_{a}} \equiv Y$ and
\begin{equation}
\tilde{\mathcal{W}}^{D}_{\alpha\beta}=\frac{2B^{L}}{\lambda^{2}_{l}}
\left[(YY^{\dagger})_{\alpha\beta}\frac{\Delta_{\Delta}}{\Sigma_{\Delta}}+
\frac{1}{4Y^{eq}_{L}}\left(2Y(\Delta^{0}_{l})^{T}Y^{\dagger}+YY^{\dagger}\Delta^{0}_{l}+
\Delta^{0}_{l}YY^{\dagger}\right)_{\alpha\beta}+
\frac{1}{2Y^{eq}_{L}}Y_{\alpha\tau}Y^{*}_{\beta\tau}\Delta_{l_{\tau}}
\right]\gamma_{D},
\end{equation}
\begin{equation}
\tilde{\mathcal{W}}^{D}_{\Delta}=Tr(\tilde{\mathcal{W}}^{D})+\tilde{\mathcal{W}}^{D}_{\tau}-W^{D}_{\phi},
\end{equation}
\begin{equation}
\tilde{\mathcal{W}}^{D}_{\tau}=
\frac{2B^{L}}{\lambda^{2}_{l}}\left[\left(YY^{\dagger}\right)_{\tau\tau}
\frac{\Delta_{\Delta}}{\Sigma^{eq}_{\Delta}}+\frac{1}{2Y^{eq}_{L}}
\left(\left(Y(\Delta^{0}_{l})^{T}Y^{\dagger}\right)_{\tau\tau}+
\left(\left(YY^{\dagger}\right)_{\tau\tau}+|Y_{\tau\tau}|^{2}\right)
\Delta_{l_{\tau}}\right)\right]\gamma_{D},
\end{equation}
and 
\begin{equation}
 W^{D}_{\phi}=2B^{\phi}\left(\frac{\Delta_{\phi}}{Y^{eq}_{\phi}}
 -\frac{\Delta_{\Delta}}{\Sigma^{eq}_{\Delta}}\right)\gamma_{D},
\end{equation}
where $(\Delta^{0}_{l})_{\alpha\beta}$, $\Delta_{l_{\tau}}$ and $\Delta_{\phi}$ are given by equations (\ref{eqn-Dl1} - \ref{eqn-Dphi}). 

Solving the flavor-covariant Boltzmann equations numerically, we obtain flavored
lepton asymmetries and we estimate the final baryon asymmetry as \cite{Lavignac:2015gpa}
\begin{equation}
 \eta_{B}=7.04\times\frac{12}{37}\times\sum_{\alpha}\Delta_{\alpha},
 \label{eqn-etaBDME}
\end{equation}
where $\Delta_{\alpha}=\Delta_{B/3-L_{\alpha}}$.
\section{Baryogenesis through flavored leptogenesis: Results}
\label{sec:results}
 We obtain the final baryon asymmetry given in Eq.(\ref{eqn-etaBDME})
via lepton asymmetries $\Delta_{\alpha}$ by solving the full set of
DMEs given in equations (\ref{eqn-DME1} - \ref{eqn-DME3}),
for two cases. The cases depend upon the mass hierarchy among the two triplet scalars;
(I) $M_{1} \ll M_2$ (II) $M_{2} \ll M_1$. In thermal leptogenesis, the Yukawa couplings 
of the corresponding triplet scalars in both cases
 play an important role in the generation of adequate CP asymmetries
and washout effects. 
We further study, in each case, the role of hierarchical branching ratios $(B^{L}_{a}\sim 0.0005,0.005,0.05)$ in generating adequate baryon asymmetry as compared to comparable branching ratios $(B^{L}_{a}\sim 0.5)$. The CP-violating phase $\alpha / \beta$
plays an important role in compensating the suppression in the CP asymmetry, which can
manifest due to hierarchy in the branching ratios. We also take different sets of triplet $\Delta_{a}$ masses $M_{a}\sim 10^{10}$ GeV, flavored CP asymmetries $\epsilon^{\alpha\beta}_{a}$ and calculate the baryon asymmetry for different
hierarchy of branching ratios.
\subsection{Formulation of flavored CP asymmetries}
 The expression of flavored CP asymmetry from triplet $\Delta_{1}$ decay
 \begin{equation}
 \epsilon^{\alpha\beta}_{1}=c_{\alpha\beta}\frac{M^{2}_{1}
 |u_{1}|^{2}}{2\pi\left(3z^{2}_{1}v^{4}+4M^{4}_{1}|u_{1}|^{4}\right)}\times
 \rm{Im}\left[m^{(1)}_{\nu,\alpha\beta}m^{*}_{\nu,\alpha\beta}\right],
 \label{eqn-epsmat}
\end{equation}
and the same originating from triplet $\Delta_{2}$ decay
\begin{equation}
 \epsilon^{\alpha\beta}_{2}=c_{\alpha\beta}\frac{M^{2}_{2}
 |u_{2}|^{2}}{4\pi\left(z^{2}_{2}v^{4}(1-2\varepsilon_{1}-2\varepsilon_{2})+2M^{4}_{2}|u_{2}|^{4}\right)}\times
 \rm{Im}\left[m^{(2)}_{\nu,\alpha\beta}m^{*}_{\nu,\alpha\beta}\right],
 \label{eqn-epsmat2}
\end{equation}
get generated from Eq. (\ref{eqn-vflav}) using the expressions of branching ratios (Eq. (\ref{eqn-Blhfunc})) and total triplet decay width
(Eq. (\ref{eqn-gmD})). 
The total neutrino mass matrix expression (Eq. (\ref{eqn-mnudiag})) is obtained by taking the modified diagonalising matrix $U$  given in Eq. (\ref{eqn-pTBM}). The $\gamma_{1,2}$, $\beta$ phase relations are taken from Eq. (\ref{eqn-gm12}). 
Using the neutrino mass matrix relation 
\begin{equation}
 m_{\nu}=U^{*}d_{\nu}U^{\dagger}=U^{*}d^{(1)}_{\nu}U^{\dagger}+U^{*}d^{(2)}_{\nu}U^{\dagger},
 \label{eqn-mnuCP}
\end{equation}
we can obtain the neutrino mass matrix contribution from triplet $\Delta_{1}$, as
\begin{equation}
 m^{(1)}_{\nu}=U^{*}d^{(1)}_{\nu}U^{\dagger}=z_{1}e^{i\beta}
 \begin{pmatrix}
  1 & 2\varepsilon_{1} & 2\varepsilon_{2}\\
  2\varepsilon_{1} & 2\varepsilon_{2} & 1\\
  2\varepsilon_{2} & 1 & 2\varepsilon_{1}
 \end{pmatrix}
 \label{eqn-mnu1}
\end{equation}
where
\begin{equation}
 d^{(1)}_{\nu}=\rm diag\left(z_{1}e^{i(\beta-\sigma_{1})},z_{1},-z_{1}e^{i(\beta-\sigma_{2})}\right).
\end{equation}
and we can similarly obtain the neutrino mass matrix contribution from triplet $\Delta_{2}$, as
\begin{equation}
 m^{(2)}_{\nu}=U^{*}d^{(2)}_{\nu}U^{\dagger}=\frac{1}{3}z_{2}(1-\varepsilon_{1}-\varepsilon_{2})
 \begin{pmatrix}
  2 & -1 & -1\\
  -1 & 2 & -1\\
  -1 & -1 & 2
 \end{pmatrix}
 \label{eqn-mnu2}
\end{equation}
\begin{equation}
 d^{(2)}_{\nu}=\rm diag\left(z_{2},0,z_{2}\right).
\end{equation}
Within the energy range $[10^{9},10^{12}]$ GeV, the flavored CP asymmetries can be expressed as $\epsilon^{\alpha \beta}_{a}(\alpha, \beta =e,\mu)$ and $\epsilon^{\tau\tau}_{a}$, where $a=1,2$ is associated with triplet $\Delta_{1,2}$. Using equations (\ref{eqn-mnuCP}), (\ref{eqn-mnu1}) and (\ref{eqn-mnu2}) we calculate, the flavored CP asymmetries. The specific expressions are given  in equations (\ref{eqn-CP1ee} - \ref{eqn-CP1tt}) and (\ref{eqn-CP2ee} - \ref{eqn-CP2tt}) in Appendix (\ref{app:epsilons}).  The dependence of $\beta$ 
on the CP asymmetries are plotted in the figures (\ref{fig-BD1}) and (\ref{fig-BD2}) for different values $B^L_{1(2)} =
0.5, 0.05, 0.005, 0.0005$.
The triplet VEV $|u_{1}|$ ($|u_{2}|$) are taken in the range $[0.1-100]$ eV. 
To estimate the CP asymmetry from the model, we have chosen the small perturbation parameters 
from the allowed parameter space shown in Fig.(\ref{fig-caseAmod}), as
$\varepsilon_{1}\sim -0.125$ and $\varepsilon_{2}\sim 0.075$.
 We have  varied 
the phase $\beta$ within the range $(100^{\circ} - 260^{\circ})$ so that it  remains consistent with the latest bound on the effective 
neutrino mass $|m_{ee}| \lesssim 0.15$ eV.
The phase $\sigma_{1}$ is taken in such a 
way that it satisfies the relations given in Eq.(\ref{eqn-gm12}). The viable 
values of the parameters $z_{1}$ and $z_{2}$ are calculated as functions of 
two mass squared differences and phase $\beta$, from the expressions 
equations (\ref{eqn-calz1}) and (\ref{eqn-calz2}).
It is observed that CP asymmetries of the order $\mathcal{O}(10^{-7})$ can be
produced for triplet masses $\sim 10^{10}$ GeV.
The figures (\ref{fig-BD1}) and (\ref{fig-BD2}) are generated by taking $M_{1}$ ($M_{2}$) $=2\times 10^{10}$ GeV and taking the mass squared differences in $3 \sigma$ range. The scale is suitable for studying two-flavor leptogenesis which we take up
in the upcoming sections.
 \subsection{Baryogenesis from $\Delta_{1}$ decay}
 First, we consider the case -I, $M_{1}\ll M_{2}$, where the lepton asymmetry is generated by the decay processes of triplet $\Delta_{1}$. We obtain the final baryon asymmetry given in Eq. (\ref{eqn-etaBDME})
via lepton asymmetries $\Delta_{\alpha}$ by solving the set of
DMEs given in equations (\ref{eqn-DME1} - \ref{eqn-DME3}),
for different sets of triplet mass $M_{1}$ and  the flavored CP asymmetries
$\epsilon^{\alpha\beta}_{1}$. The results of baryon asymmetry for some of the benchmark points are briefed in table (\ref{table-DMEres1}).
 In table (\ref{table-DMEres1}), we show the interplay of hierarchical branching ratios and the SCPV phase $\beta$ on the flavored CP asymmetry parameters. For example, for 
$M_1 = 3\times 10^{10}$ GeV, using Eq. (\ref{eq:BR-L}), we calculate the branching ratios,
$B_1^L$ using the allowed values of the parameters of the model. Using the same values
of the parameters we also calculate the corresponding  CP asymmetries as shown in the table. While the hierarchy in $B_1^L$ increases, it should have resulted in suppression
of the magnitude of the CP asymmetries (as can be seen in Eq. (\ref{eqn-vflav})).
But the contribution of the SCPV phase $\beta$ in $\epsilon^{\alpha\beta}_{1}$ (equations 
(\ref{eqn-CP1ee} - \ref{eqn-CP1tt})) compensates the suppression arising from the
increase in the hierarchy of the branching ratios. It should be noted here that $B_1^L$
is independent of the phase $\beta$. This feature also enhances the efficiency of
leptogenesis which can be seen from the increase in the final baryon asymmetry as shown in the table.
 \begin{table}[h!]
 \centering 
 \begin{tabular}{|p{1.5cm}||p{1.2cm}||p{1.2cm}||p{1.5cm}||p{1.5cm}||p{1.5cm}||p{1.5cm}||p{1.8cm}|}
 \hline
$M_{1}({\rm GeV})$&$B^{L}_{1}$&$\beta (^{\circ})$&$\epsilon^{ee}_{1}/10^{-8}$&$\epsilon^{e\mu}_{1}/10^{-9}$&$\epsilon^{\mu \mu}_{1}/10^{-9}$&$\epsilon^{\tau \tau}_{1}/10^{-9}$&$|\eta_{B}|/10^{-10}$\\
\hline
 &$0.48$&$178$&$0.70$&$0.84$&$1.01$&$-1.68$&$0.04$\\
 $3\times 10^{10}$&$0.06$&$175$&$2.06$&$2.46$&$2.94$&$-4.91$&$0.29$\\
 &$0.005$&$165$&$2.04$&$2.43$&$2.91$&$-4.86$&$0.73$\\
 &$0.0005$&$126$&$2.06$&$2.46$&$2.94$&$-4.91$&$1.74$\\
 \hline
 &$0.34$&$178$&$2.79$&$3.32$&$3.98$&$-6.63$&$0.30$\\
 $4\times 10^{10}$&$0.05$&$175$&$2.82$&$3.36$&$4.03$&$-6.71$&$0.44$\\
 &$0.005$&$164$&$2.82$&$3.36$&$4.03$&$-6.71$&$1.07$\\
 &$0.0006$&$126$&$2.83$&$3.37$&$4.04$&$-6.74$&$2.50$\\
 \hline
\end{tabular}
\caption{Baryon asymmetries through flavored leptogenesis for the triplet mass hierarchy case-1, $M_{1}\ll M_{2}$ by solving DME in the energy range
$[10^{9},10^{12}]$ GeV for different values of $M_{1}\sim 10^{10}$ GeV, $\epsilon^{\alpha\beta}_{1}$, and hierarchical $(B^{L}_{1}\sim 0.0005,0.005,0.05)$ as well as comparable $(B^{L}_{1}\sim 0.5)$ branching ratio $B^{L}_{1}$.}  
\label{table-DMEres1}
\end{table}

\begin{figure*} 
 \centering
\begin{subfigure}{0.45\linewidth}
 \includegraphics[width=\linewidth]{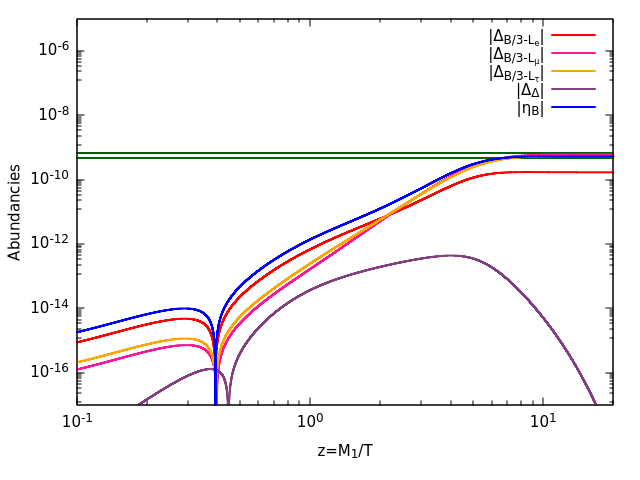}
\caption{}
\label{fig-set1DME}
\end{subfigure}
 \begin{subfigure}{0.45\linewidth}
\includegraphics[width=\linewidth]{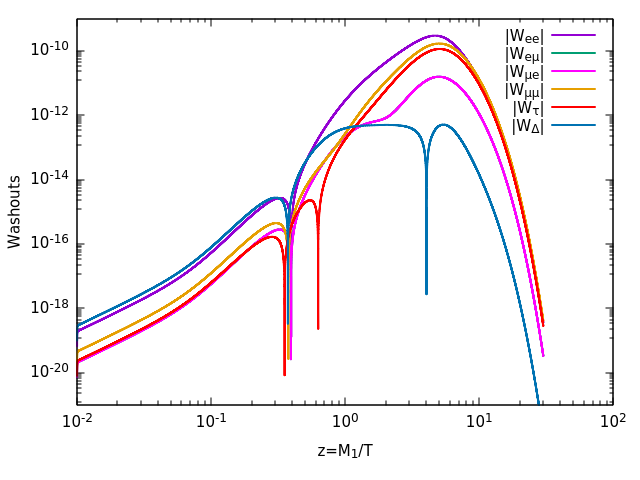}
\caption{}
\label{fig-set1DME1}
\end{subfigure}
\caption{Solution of DMEs for asymmetries and washouts for set with $M_{1}=3\times 10^{10}$ GeV
by using the flavored CP asymmetries from table (\ref{table-D1setres}). Left: The color lines
represent different asymmetries. Using the numerically obtained asymmetries
the final baryon asymmetry is calculated. The green lines correspond to the
observational value of the BAU.
Right: The color lines represent numerically obtained amount of washouts along different flavor directions in units of $sHz$.}
\label{fig-D1set}
\end{figure*}

 We present a specific result corresponding to a set with $M_{1}=3\times 10^{10}$ GeV, and flavored CP asymmetries $\epsilon^{ee}_{1}\sim 10^{-7}$, $\epsilon^{e\mu,~\mu e,~\mu\mu,~\tau\tau}_{1}\sim 10^{-8}$ as given in table (\ref{table-D1setres}). We have plotted the obtained flavored lepton asymmetries and baryon asymmetry in Fig. (\ref{fig-set1DME}) along with the evolution of different washout terms in  Fig. (\ref{fig-set1DME1}).  
 Although we have shown the results for final baryon asymmetry for a few benchmark points,
we have taken different data sets corresponding to particular flavored CP asymmetries 
for $M_{1}\sim10^{10}$ GeV and hierarchical branching ratios and we have plotted the obtained baryon asymmetry for each set in Fig. (\ref{fig-set1DMEa}).
\begin{table}[h!]
 \centering 
 \begin{tabular}{|p{1.6cm}||p{0.9cm}||p{1.4cm}||p{1.5cm}||p{1.5cm}||p{1.5cm}||p{1.5cm}||p{2.0cm}|}
 \hline
$M_{1}({\rm GeV})$&$B^{L}_{1}$&$|u_{1}|(\rm eV)$&$\epsilon^{ee}_{1}/10^{-7}$&$\epsilon^{e\mu}_{1}/10^{-8}$&$\epsilon^{\mu\mu}_{1}/10^{-8}$&$\epsilon^{\tau\tau}_{1}/10^{-8}$&$|\eta_{B}|/10^{-10}$\\
\hline
$3\times 10^{10}$ Fig. (\ref{fig-D1set}) &$0.58$&$4.80$&$4.47$&$5.32$&$6.38$&$-1.06$&$5.29$\\
\hline
\end{tabular}
\caption{Baryon asymmetries through leptogenesis by solving DME in the energy range
$[10^{9},10^{12}]$ GeV for $M_{1}=3\times 10^{10}$ GeV.}  
\label{table-D1setres}
\end{table}
\subsection{Baryogenesis from $\Delta_{2}$ decay}
\begin{figure*} 
 \centering
\begin{subfigure}{0.45\linewidth}
 \includegraphics[width=\linewidth]{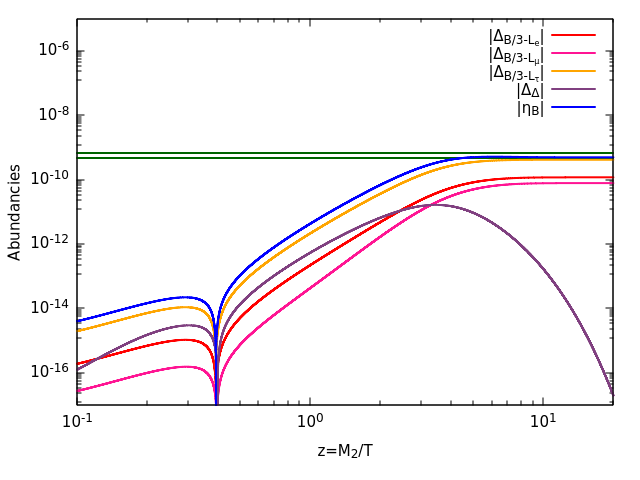}
\caption{}
\label{fig-set2DME}
\end{subfigure}
 \begin{subfigure}{0.45\linewidth}
\includegraphics[width=\linewidth]{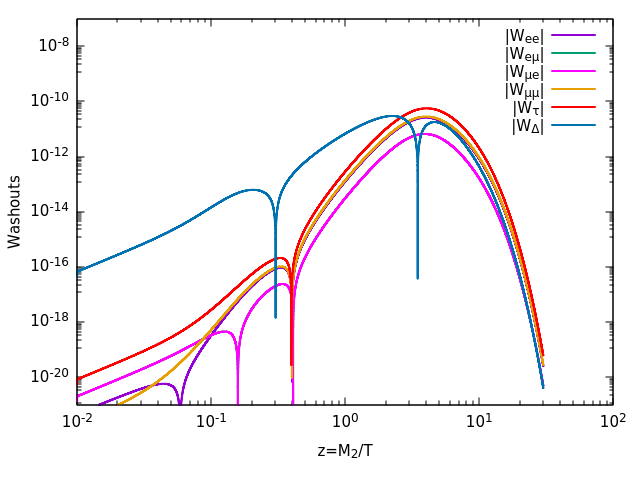}
\caption{}
\label{fig-set2DME1}
\end{subfigure}
\caption{Solution of DMEs for asymmetries and washouts for set with $M_{2}=1.4\times 10^{10}$ GeV
by using the flavored CP asymmetries from table (\ref{table-D2setres}). Left: The color lines
represent different asymmetries. Using the numerically obtained asymmetries
the final baryon asymmetry is calculated. The green lines correspond to the
observational value of the BAU.
Right: The color lines represent numerically obtained amount of washouts along different flavor directions in units of $sHz$.}
\label{fig-D2set}
\end{figure*}

\begin{figure} 
 \centering
\begin{subfigure}{0.45\linewidth}
 \includegraphics[width=\linewidth]{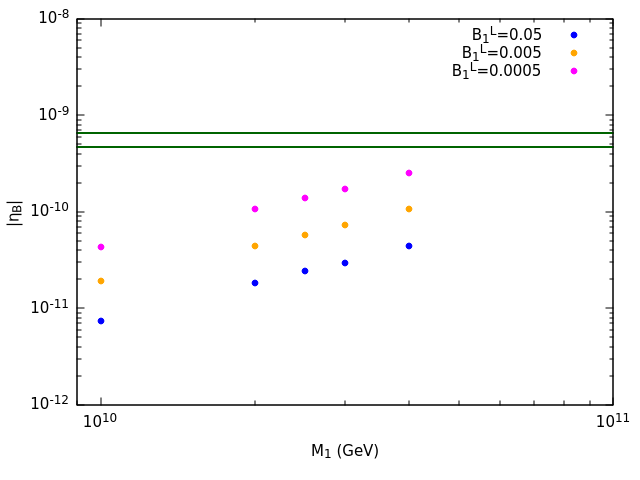}
\caption{}
\label{fig-set1DMEa}
\end{subfigure}
 \begin{subfigure}{0.45\linewidth}
\includegraphics[width=\linewidth]{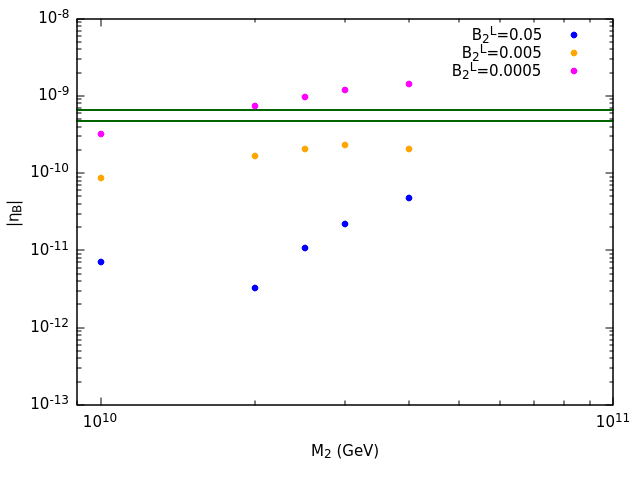}
\caption{}
\label{fig-set1DMEb}
\end{subfigure}
\caption{Plot of obtained baryon asymmetries $|\eta_{B}|$ for different masses 
of the triplet scalars in left (right) for $M_{1}(M_2)$ in GeV. The figures also show a comparative study of the efficiency of leptogenesis and its dependence on the branching ratio hierarchy. The bound in green shows the range of the baryon asymmetry
from observations.}
\label{fig-rangesDME1}
\end{figure}

We further consider the triplet mass hierarchy case (II), with $M_{2}\ll M_{1}$ and take different sets with triplet $\Delta_{2}$ mass $M_{2}\sim 10^{10}$ GeV, flavored CP asymmetries and hierarchical branching ratios. The results are similar to case (I).
The results are briefed in table (\ref{table-DMEres2}). In this case, for comparatively
lower values of triplet mass, the model can result in adequate baryon asymmetry.
We present a specific result corresponding to a set with $M_{2}=1.4 \times 10^{10}$ GeV, and flavored CP asymmetries $\epsilon^{ee}_{2}\sim 10^{-7}$, $\epsilon^{e\mu,~\mu e,~\mu\mu}_{2}\sim 10^{-8}$, $\epsilon^{\tau\tau}_{2}\sim 10^{-6}$ as given in 
table (\ref{table-D2setres}). We have plotted the obtained flavored lepton asymmetries and baryon asymmetry in Fig. (\ref{fig-set2DME}) along with the evolution of different washout terms in Fig. (\ref{fig-set2DME1}).  
Similar to the case -I,
we have taken different data sets corresponding to particular flavored CP asymmetries 
for $M_{2}\sim10^{10}$ GeV and hierarchical branching ratios and we have plotted the obtained baryon asymmetry for each set in Fig. (\ref{fig-set1DMEb}).
\begin{table}[htb!]
 \centering 
 \begin{tabular}{|p{1.5cm}||p{1.2cm}||p{1.2cm}||p{1.5cm}||p{1.5cm}||p{1.5cm}||p{1.5cm}||p{1.8cm}|}
 \hline
$M_{2}({\rm GeV})$&$B^{L}_{2}$&$\beta (^{\circ})$&$\epsilon^{ee}_{2}/10^{-8}$&$\epsilon^{e\mu}_{2}/10^{-9}$&$\epsilon^{\mu \mu}_{2}/10^{-9}$&$\epsilon^{\tau \tau}_{2}/10^{-7}$&$|\eta_{B}|/10^{-10}$\\
\hline
&$0.59$&$176$&$-0.66$&$-0.79$&$-0.95$&$0.75$&$0.04$\\
$1\times 10^{10}$&$0.06$&$172$&$-0.67$&$-0.80$&$-0.96$&$0.76$&$0.07$\\
&$0.005$&$157$&$-0.67$&$-0.80$&$-0.96$&$0.75$&$0.86$\\
&$0.0006$&$126$&$-0.73$&$-0.87$&$-1.05$&$0.75$&$3.26$\\
\hline
&$0.53$&$177$&$-1.29$&$-1.53$&$-1.84$&$1.45$&$0.01$\\
$2\times 10^{10}$&$0.05$&$176$&$-1.42$&$-1.68$&$-2.02$&$1.59$&$0.03$\\
&$0.006$&$156$&$-1.42$&$-1.69$&$-2.03$&$1.58$&$1.66$\\
&$0.0005$&$126$&$-1.42$&$-1.69$&$-2.03$&$1.45$&$7.51$\\
 \hline
\end{tabular}
\caption{Baryon asymmetries through flavored leptogenesis for the triplet mass hierarchy case - II, $M_{2}\ll M_{1}$ by solving DME in the energy range
$[10^{9},10^{12}]$ GeV for different values of $M_{2}\sim 10^{10}$ GeV, $\epsilon^{\alpha\beta}_{2}$, and hierarchical $(B^{L}_{2}\sim 0.0005,0.005,0.05)$ as well as comparable $(B^{L}_{2}\sim 0.5)$ branching ratio $B^{L}_{2}$.}  
\label{table-DMEres2}
\end{table}
\begin{table}[htb!]
 \centering 
 \begin{tabular}{|p{1.8cm}||p{0.9cm}||p{1.4cm}||p{1.5cm}||p{1.5cm}||p{1.5cm}||p{1.5cm}||p{1.8cm}|}
 \hline
$M_{2}({\rm GeV})$&$B^{L}_{2}$&$|u_{2}|(\rm eV)$&$\epsilon^{ee}_{2}/10^{-8}$&$\epsilon^{e\mu}_{2}/10^{-9}$&$\epsilon^{\mu \mu}_{2}/10^{-9}$&$\epsilon^{\tau \tau}_{2}/10^{-7}$&$|\eta_{B}|/10^{-10}$\\
\hline
$1.4\times 10^{10}$ Fig. (\ref{fig-D2set}) &$0.0006$&$48.1$&$-1.00$&$-1.20$&$-1.44$&$1.04$&$4.92$\\
\hline
\end{tabular}
\caption{Baryon asymmetries through leptogenesis by solving DME in the energy range
$[10^{9},10^{12}]$ GeV for $M_{2}=1.4\times 10^{10}$ GeV.}  
\label{table-D2setres}
\end{table}
\subsection{Summary}
\label{sec:disc}
The results obtained in this section can be summarized as follows:
\begin{enumerate}
 \item The SCPV phase $\beta$ acts as a common link between the high energy CPV in leptogenesis and low energy effect such as neutrinoless double beta decay as shown in Fig. (\ref{fig-meeAmod1}).
\item The hierarchical branching ratios result in different amounts of washout effects
 along different flavor directions. The amount of washouts greatly influences the efficiency of leptogenesis. Comparing figures (\ref{fig-set1DME1}) ($B_1^L = 0.58$) and (\ref{fig-set2DME1}) ($B_2^L = 0.0006$), one can see that in the first case, the washout effects along the flavor directions are more as compared to the second case (except $ee$-flavor direction).
This feature can be attributed to the specific flavor
 structure of the triplet Yukawa coupling matrices of the model. Although, in the second case, the extreme hierarchy in the branching ratios can in principle suppress 
 the flavored CP asymmetry parameters, the presence of SCPV phase $\beta$ can neutralise
 the effect. 
 \item The process of generation of baryon asymmetry from lepton asymmetry requires 
 that CP violation should be maximal. Observing Eq. (\ref{eqn-vflav}), it is ideal to
 generate maximal CP asymmetry for lower values of $M_1(M_2)$. This would require 
 $B^L \sim B^\phi$. In our analysis, in bringing down the $M_1(M_2)$ scale as low as possible it can be observed that
\begin{enumerate}
\item 
In case - I, $M_{1}\ll M_{2}$, (in Fig. (\ref{fig-D1set}), also Fig. (\ref{fig-set1DMEa})) the observable range of baryon asymmetry can be achieved for triplet mass $M_{1}$ as low as $\sim 3\times 10^{10}$ GeV for comparable branching ratios $B^{L}_{1}\sim 0.5$. Taking hierarchical branching ratios and requiring adequate flavored CP asymmetries,
baryon asymmetry can only be produced in the observable range for a triplet mass $M_{1} >  5\times 10^{10}$ GeV for a highly hierarchical branching ratio $B^{L}_{1}\sim0.0005$.
\item In case - II, $M_{2}\ll M_{1}$, we find from Fig. \ref{fig-D2set} 
( also Fig. (\ref{fig-set1DMEb})) the observational range of baryon asymmetry can be achieved for triplet mass $M_{2}$ as low as $\sim 1\times 10^{10}$ GeV for highly hierarchical branching ratio $B^{L}_{2}\sim 0.0005$. 

\item From Fig. (\ref{fig-rangesDME1}), for both cases - I and II, with hierarchical branching ratios the baryon asymmetry increases for a fixed value of $M_1(M_2)$. 
As shown in tables (\ref{table-DMEres1}) and (\ref{table-DMEres2}), the suppression in
CP asymmetry due to the hierarchical branching ratio, can be compensated by the 
extra source of CP violation due to the presence  of the SCPV phase $\beta$
in the neutrino mass matrix. As a result,  
even with the modified washout effect from hierarchical branching ratios, it is 
possible to achieve maximal CP violation and 
baryon asymmetry  in the observable range. The flavor structure of the
model is suitable to draw these conclusions.
\end{enumerate}
\end{enumerate}
\section{Conclusion}
\label{sect-conc}
Spontaneous violation of CP symmetry is very promising in 
explaining CP violation in low-energy processes like neutrino oscillation and high-energy
process like leptogenesis. In general CP violation in these two sectors are 
not related \cite{Davidson:2007va} and are model-dependent.
In this paper, we have studied, in an extended Standard model, neutrino mass generated via type-II seesaw mechanism and leptogenesis from the decay of heavy scalar triplet with a common source of CP violation. 
The model  provides a viable explanation of baryon asymmetry through
thermal leptogenesis, by sourcing enough CP symmetry spontaneously 
broken due to the presence of a scalar singlet with complex VEV. 
Moreover, the neutrino mass matrix exhibits TBM structure which 
generally fails to address the occurrence of non-zero CHOOZ mixing
angle, $\theta_{13}$.  By incorporating small perturbations around flavon VEV in
the high-energy scale, it can accommodate $\theta_{13}\neq0$ 
 in the model. 
With the latest neutrino oscillation data and bound on 
 the effective neutrino mass, we have analyzed the 
 parameter space of the model. Using the constrained parameter space of the model 
 we have studied flavored leptogenesis from the decay of the triplet scalars.
 The flavor structure of the model suppresses 
 flavor-independent CP asymmetry and hence we explore 
 the effect of flavor in leptogenesis by using a set of 
 flavor-covariant Boltzmann equations considering a density 
 matrix formalism. In our study, the scale of leptogenesis
 is $\mathcal{O}(10^{10})$ GeV. We also study the role of
 hierarchical branching ratios and the SCPV phase $\beta$ in generating
 maximal CP asymmetry to generate baryon asymmetry
 in the observational range. Requiring mass of triplet as
 low as possible, we find BAU in the
 observable range can be produced for $M_1 = 3\times 10^{10}$ GeV
 for comparable branching ratios. Whereas the same can be achieved for 
 $M_2 =1.4 \times 10^{10}$ GeV for hierarchical branching ratios. The flavor structure of 
 the triplet Yukawa couplings plays an important role 
 in generating the washout effects which are inevitable for
 thermal leptogenesis. 

\appendix
\section{Scalar potential of the model}
\label{App:A}
The scalar potential invariant under the symmetries as given in table (\ref{table-fieldsrep}),
can be generally realised as
\begin{equation}
 V=V_{S}+V_{\phi}+V_{\Delta}+V_{S\phi}+V_{S\Delta}+V_{\phi\Delta}+V_{S\phi\Delta},
\end{equation}
where
\begin{flalign}
 V_{S} &=\mu^{2}_{S}(S^{2}+S^{*2})+m^{2}_{S}S^{*}S+
 \lambda'_{S}(S^{4}+S^{*4})+\lambda''_{S}S^{*}S(S^{2}+S^{*2})+\lambda_{S}(S^{*}S)^{2},&
\end{flalign}
 \begin{flalign}
 V_{\phi}&=m^{2}_{\phi}\phi^{\dagger}\phi+\lambda_{\phi}(\phi^{\dagger}\phi)^{2},&
\end{flalign}
 \begin{flalign}
 V_{\Delta}&=\Sigma_{a}M^{2}_{a}Tr(\Delta^{\dagger}_{a}\Delta_{a})
 +\Sigma_{a,b}[(\lambda_{\Delta})_{ab}Tr(\Delta^{\dagger}_{a}\Delta_{a})Tr(\Delta^{\dagger}_{b}\Delta_{b})
 +(\lambda'_{\Delta})_{ab}Tr(\Delta^{\dagger}_{a}\Delta_{a}\Delta^{\dagger}_{b}\Delta_{b})],&
\end{flalign}

\begin{flalign}
V_{S\phi}&=\eta_{S}(S^{*}S)(\phi^{\dagger}\phi)+\eta'_{S}(S^{2}+S^{*2})(\phi^{\dagger}\phi),&
\end{flalign}

 \begin{flalign}
 V_{S\Delta}&=\Sigma_{a}Tr(\Delta^{\dagger}_{a}\Delta_{a})[\eta_{a}(S{2}+S^{*2})+\xi_{a}S^{*}S],&
\end{flalign}

\begin{flalign}
 V_{\phi\Delta}&=\Sigma_{a}[\xi'_{a}(\phi^{\dagger}\phi)Tr(\Delta^{\dagger}_{a}\Delta_{a})
 +\xi''_{a}(\phi^{\dagger}\Delta^{\dagger}_{a}\Delta_{a}\phi)]+(\mu_{2}M_{2}\tilde{\phi}^{T}\Delta_{2}\tilde{\phi}+{\rm H.c.}),&
 \end{flalign}

 \begin{flalign}
 V_{S\phi\Delta}&=\tilde{\phi}^{T}\Delta_{1}\tilde{\phi}(\lambda_{1}S+\lambda'_{1}S^{*})+ {\rm H.c.},&
 \label{eqn-S}
\end{flalign}
where $\tilde{\phi}=i\sigma_{2}\phi^{*}$.
It is interesting to note that a viable leptogenesis scenario in this context requires 
a term $\mu_{1}M_{1}\tilde{\phi}^{T}\Delta_{1}\tilde{\phi}$ such a term is forbidden due 
to the choice of $Z_{4}$ symmetry. However, the complex VEV acquired by the scalar singlet
$S$ enables this term to be generated from the scalar potential as shown in Eq. (\ref{eqn-S}).
Furthermore, the addition of two heavy scalar fields allows us to explore the spontaneous 
breaking of the discrete $A_{4}$ symmetry.  

The potential can be written in tree level as 
\begin{equation}
 V_{0}=m^{2}_{S}v^{2}_{S}+\lambda_{S}v^{4}_{S}+2(\mu^{2}_{S}
 +\lambda''_{S}v^{2}_{S})v^{2}_{S}\cos(2\alpha)+2\lambda'_{S}v^{4}_{S}\cos(4\alpha).
\end{equation}
Minimizing the potential with respect to $v_{S}$ and $\alpha$ we obtain,
\begin{equation}
 \frac{\partial V_{0}}{\partial v_{S}}=2v_{S}[m^{2}_{S}+2\lambda_{S}v^{2}_{S}+2(\mu^{2}_{S}+2\lambda''_{S}v^{2}_{S})\cos(2\alpha)+4\lambda'_{S}v^{2}_{S}\cos(4\alpha)]=0,
\end{equation}
and
\begin{equation}
 \frac{\partial V_{0}}{\partial \alpha}=-4v^{2}_{S}\sin(2\alpha)[(\mu^{2}_{S}+\lambda''_{S}v^{2}_{S})+4\lambda'_{S}v^{2}_{S}\cos(2\alpha)]=0.
\end{equation}
Three possible solution to the above equations, excluding the trivial solution $v_{S}=0$ that leads to $V_{0}=0$, are given by 
\begin{enumerate}
\item 
\begin{equation}
v^{2}_{S}=-\frac{m^{2}_{S}+2\mu^{2}_{S}}{2(\lambda_{S}+2\lambda'_{S}+2\lambda''_{S})}, \quad \alpha=0,\pm \pi,
\end{equation}
\item 
\begin{equation}
v^{2}_{S}=\frac{-m^{2}_{S}+2\mu^{2}_{S}}{2(\lambda_{S}+2\lambda'_{S}-2\lambda''_{S})},\quad \alpha=\pm\frac{\pi}{2},
\label{eqn-vs2}
\end{equation}
\item 
\begin{equation}
 v^{2}_{S}=\frac{-2\lambda'_{S}m^{2}_{S}+\lambda''_{S}\mu^{2}_{S}}{4\lambda_{S}\lambda'_{S}-
 8\lambda'_{S}-\lambda''_{S}},\quad \cos(2\alpha)=-\frac{\mu^{2}_{S}
 +\lambda''_{S}v^{2}_{S}}{4\lambda'_{S}v^{2}_{S}}.
\end{equation}
\end{enumerate}
To show that the last solution, which also leads in this case to the spontaneous
breaking of the CP symmetry, corresponds to the global minimum of the potential, 
we consider $m^{2}_{S}<0$, $\lambda''_{S}\simeq 0$, $\mu_{S}\simeq 0$ and $\lambda_{S}>2\lambda'_{S}>0$ 
and we obtain from Eq. (\ref{eqn-vs2}),
\begin{equation}
 v^{2}_{S}\simeq -\frac{m^{2}_{S}}{2(\lambda_{S}+2\lambda'_{S})}, \quad 
 \alpha=0,\pm\frac{\pi}{2},\pm\pi.
\end{equation}
For cases - 1, and 2,
\begin{equation}
 V_{0}\simeq -\frac{m^{4}_{S}}{4(\lambda_{S}+2\lambda'_{S})}.
\end{equation}
For case - 3
\begin{equation}
 v^{2}_{S}\simeq -\frac{m^{2}_{S}}{2(\lambda_{S}-2\lambda'_{S})},\quad \alpha\simeq \pm\frac{\pi}{4},
\end{equation}
which leads to
\begin{equation}
 V_{0}\simeq -\frac{m^{4}_{S}}{4(\lambda_{S}-2\lambda'_{S})}
\end{equation}
which is the absolute minimum of the potential.
\section{Reaction Densities}
\label{app:rates}
\begin{figure}[h]
 \centering
\includegraphics[scale=0.5]{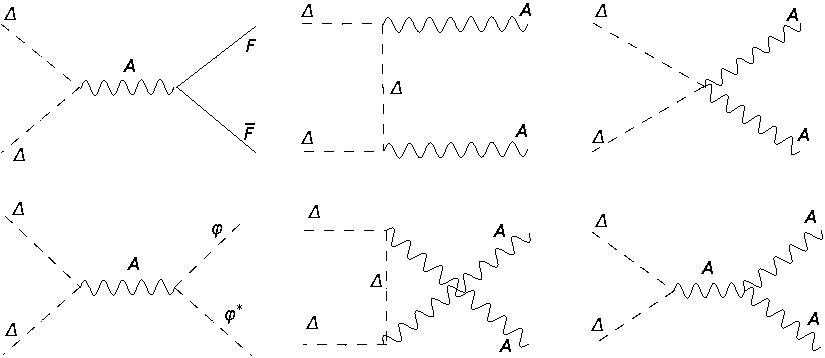}
\caption{Feynman diagrams that contribute to the interaction rate $\gamma_A$}
\label{gammaA}
\end{figure}

  The  space-time density of total decay of triplet scalar $\Delta$ and its
  anti-particle $\overline{\Delta}$, $\gamma_D$ and  that of $2\leftrightarrow2$ 
  scattering, $\gamma_A$ are  given by,
 \begin{equation}
 \gamma_D=s\Gamma_\Delta\Sigma^{eq}_\Delta\frac{K_1(z)}{K_2(z)},
 \end{equation}
 \begin{equation}
 \gamma_A =\frac{M^4_a}{64\pi^4}\int^{\infty}_{x_{\rm min}} dx  
 \sqrt{x}\frac{K_1\left( z \sqrt{x} \right) \hat{\sigma}_S}{z}.
 \label{eq:gamma-gen}
 \end{equation}
 Here, $x= {\rm s}/M^2_a$, $\textbf{s}$ is the Mandelstam variable, square
 of center of mass energy \cite{Halzen:1984mc}, $\hat{\sigma}_S$ is the 
 reduced cross-section of $2\rightarrow 2$ scatterings (relevant diagrams
 in fig.(\ref{gammaA})). For processes mediated by gauge bosons $x_{\rm min} =4$
 and for Yukawa-induced reactions $x_{\rm min} =0$. 
 
 The corresponding reaction density of scattering is given by,
 \begin{equation}
 \gamma_A=\frac{M_a T^3e^{-\frac{2M_a}{T}}}{64\pi^4}(9g^4
 +12g_2^2g^2_{\mathcal{Y}}+3g^4_Y)\left(1+\frac{3T}{4M_a}\right).
 \label{eq:gammaA}
 \end{equation}
Here we take $\Gamma_{\Delta}=\frac{M^2_a \tilde{m}_\Delta}{16\pi v^2 \sqrt{B^L B^{\phi}}}$, 
$\tilde{m}_\Delta = {\rm Tr} (m^{(a)\dagger}_\nu m^{(a)}_\nu )$, $\Sigma^{\rm eq}_\Delta=2Y^{\rm eq}_\Delta
=2\times\frac{45g_\Delta}{4\pi^4g_{*}}z^2K_2(z)$ and
$g_\Delta=3$, $g_{*}=106.75$ are the triplet scalar and SM particle-degrees of freedom,
respectively \cite{Felipe:2013kk}. 
$g_2, g_{\mathcal{Y}}$ are the coupling constants corresponding to $SU(2)_L$ and 
$U(1)_{\mathcal{Y}}$ interactions, respectively, having values,
$g_2 =\frac{e}{\sin\theta_W}=0.651742$ and $g_{\mathcal{Y}}=\frac{e}{\cos\theta_W}=0.461388$. 
The other reaction densities, $\gamma$'s are to be calculated using the integration given in
Eq. (\ref{eq:gamma-gen}).

Within the temperature range $10^{9}$ GeV $<T<10^{12}$ GeV, the chemical 
potential relations are given by:
\begin{equation}
\sum_{i=1,2,3} (\mu_{q_{i}}+2\mu_{u_{i}}-\mu_{d_{i}})-\sum_{\alpha=e, \mu, \tau}(\mu_{l_{\alpha}}+\mu_{e_{\alpha}})+2\mu_{\phi}6\mu_{\Delta}=0 \quad (\Delta_{Y}=0)
\label{eqn-chemp1}
\end{equation}
\begin{equation}
\sum_{i=1,2,3}(2\mu_{q_{i}}+\mu_{u_{i}}+\mu_{d_{i}})=0 \quad (\Delta_{B}=0)
\end{equation}
\begin{equation}
\sum_{i=1,2,3}(2\mu_{q_{i}}-\mu_{u_{i}}-\mu_{d_{i}})=0 \quad (\text{QCD sphaleron})
\end{equation}
\begin{equation}
  \sum_{i=1,2,3}2\mu_{q_{i}}+\sum_{\alpha=e, \mu, \tau}\mu_{l_{\alpha}}=0 \quad (\text{EW sphaleron})
 \end{equation}
 \begin{equation}
  \mu_{q_{i}}-\mu_{u_{i}}+\mu_{\phi}=0 \quad (\text{up-type quark Yukawa})
 \end{equation}
 \begin{equation}
  \mu_{q_{i}}-\mu_{d_{i}}-\mu_{\phi}=0 \quad (\text{down-type quark Yukawa})
\end{equation}
 \begin{equation}
  \mu_{l_{\alpha}}-\mu_{e_{\alpha}}-\mu_{\phi}=0 \quad (\text{charged lepton Yukawa})
  \label{eqn-chemp7}
 \end{equation} 
\section{Estimation of flavored CP asymmetries}
\label{app:epsilons}
The important factors appearing in the flavored CP asymmetry parameters as given in 
Eq. (\ref{eqn-vflav}) are calculated as follows:
\begin{equation}
 \rm{Im}\left[m^{(1)}_{\nu,ee}m^{*}_{\nu,ee}\right]=
 \frac{2}{3}z_{1}\left(z^{2}_{1}+z^{2}_{2}+2z_{1}z_{2}\cos\beta\right)^{\frac{1}{2}}
 \left(1-\varepsilon_{1}-\varepsilon_{2}\right)\sin(\beta-\sigma_{1}),
 \label{eqn-Im1ee}
\end{equation}
\begin{eqnarray}
\nonumber
 \rm{Im}\left[m^{(1)}_{\nu,e\mu}m^{*}_{\nu,e\mu}\right]&=&-
 \frac{2}{3}\varepsilon_{1}z_{1}\left(z^{2}_{1}+z^{2}_{2}+2z_{1}z_{2}\cos\beta\right)^{\frac{1}{2}}
 \sin(\beta-\sigma_{1})\\
 &=&
 \rm{Im}\left[m^{(1)}_{\nu,\mu e}m^{*}_{\nu,\mu e}\right],
 \label{eqn-Im1em}
\end{eqnarray}
\begin{eqnarray}
\nonumber
 \rm{Im}\left[m^{(1)}_{\nu,\mu\mu}m^{*}_{\nu,\mu\mu}\right]&=&
 \varepsilon_{2}z_{1}\Big[\frac{1}{3}\left(z^{2}_{1}+z^{2}_{2}+2z_{1}z_{2}\cos\beta\right)^{\frac{1}{2}}\sin(\beta-\sigma_{1})\\
 &+&
 \left(z^{2}_{1}+z^{2}_{2}-2z_{1}z_{2}\cos\beta\right)^{\frac{1}{2}}\sin(\beta-\sigma_{2})\Big],
 \label{eqn-Im1mm}
 \end{eqnarray}
\begin{eqnarray}
\nonumber
 \rm{Im}\left[m^{(1)}_{\nu,\tau\tau}m^{*}_{\nu,\tau\tau}\right]&=&
 \varepsilon_{1}z_{1}\Big[\frac{1}{3}\left(z^{2}_{1}+z^{2}_{2}+2z_{1}z_{2}\cos\beta\right)^{\frac{1}{2}}\sin(\beta-\sigma_{1})\\
 &+& \left(z^{2}_{1}+z^{2}_{2}-2z_{1}z_{2}\cos\beta\right)^{\frac{1}{2}}\sin(\beta-\sigma_{2})\Big],
 \label{eqn-Im1tt}
 \end{eqnarray}
\begin{eqnarray}
\nonumber
 \rm{Im}\left[m^{(2)}_{\nu,ee}m^{*}_{\nu,ee}\right]&=&
 -\frac{2}{3}z_{2}\Big[\frac{2}{3}\left(z^{2}_{1}+z^{2}_{2}+2z_{1}z_{2}\cos\beta\right)^{\frac{1}{2}}(1-2\varepsilon_{1}-2\varepsilon_{2})\sin\sigma_{1}\\
 &+&\frac{1}{3}z_{1}(1+\varepsilon_{1}+\varepsilon_{2})\sin\beta\Big],
 \label{eqn-Im2ee}
\end{eqnarray}
\begin{eqnarray}
\nonumber
 \rm{Im}\left[m^{(2)}_{\nu,e\mu}m^{*}_{\nu,e\mu}\right]&=&\frac{1}{3}z_{2}\Big[(\frac{7}{6}\varepsilon_{1}+\frac{1}{6}\varepsilon_{2}-\frac{1}{3})\left(z^{2}_{1}+z^{2}_{2}+2z_{1}z_{2}\cos\beta\right)^{\frac{1}{2}}\sin\sigma_{1}\\
 \nonumber
 &+&\frac{1}{3}(1+\varepsilon_{1}+\varepsilon_{2})z_{1}\sin\beta\\
 \nonumber
 &-&\frac{1}{2}(\varepsilon_{1}-\varepsilon_{2})\left(z^{2}_{1}+z^{2}_{2}-2z_{1}z_{2}\cos\beta\right)^{\frac{1}{2}}\sin\sigma_{2}\Big]\\
 \nonumber
 &=&
 \rm{Im}\left[m^{(2)}_{\nu,\mu e}m^{*}_{\nu,\mu e}\right],
 \label{eqn-Im2em}
\end{eqnarray}
\begin{eqnarray}
\nonumber
 \rm{Im}\left[m^{(2)}_{\nu,\mu\mu}m^{*}_{\nu,\mu\mu}\right]&=&
-\frac{2}{3}z_{2}\Big[(-\frac{5}{6}\varepsilon_{1}+\frac{1}{6}\varepsilon_{2}-\frac{1}{6})\left(z^{2}_{1}+z^{2}_{2}+2z_{1}z_{2}\cos\beta\right)^{\frac{1}{2}}\sin\sigma_{1}\\
\nonumber
 &+&\frac{1}{3}(1+\varepsilon_{1}+\varepsilon_{2})z_{1}\sin\beta\\
 &+&\frac{1}{2}(1-\varepsilon_{1}-3\varepsilon_{2})\left(z^{2}_{1}+z^{2}_{2}-2z_{1}z_{2}\cos\beta\right)^{\frac{1}{2}}\sin\sigma_{2}\Big],
 \label{eqn-Im2mm}
\end{eqnarray}
and 
\begin{eqnarray}
\nonumber
 \rm{Im}\left[m^{(2)}_{\nu,\tau\tau}m^{*}_{\nu,\tau\tau}\right]&=&
 -\frac{2}{3}z_{2}\Big[(\frac{1}{6}\varepsilon_{1}-\frac{5}{6}\varepsilon_{2}+\frac{1}{6})\left(z^{2}_{1}+z^{2}_{2}+2z_{1}z_{2}\cos\beta\right)^{\frac{1}{2}}\sin\sigma_{1}\\
 \nonumber
 &+&\frac{1}{3}(1+\varepsilon_{1}+\varepsilon_{2})z_{1}\sin\beta\\
 &+&\frac{1}{2}(1-3\varepsilon_{1}-\varepsilon_{2})\left(z^{2}_{1}+z^{2}_{2}-2z_{1}z_{2}\cos\beta\right)^{\frac{1}{2}}\sin\sigma_{2}\Big].
 \label{eqn-Im2tt}
\end{eqnarray}

Finally from equations (\ref{eqn-epsmat}), (\ref{eqn-Im1ee}), (\ref{eqn-Im1em}), (\ref{eqn-Im1mm}) and (\ref{eqn-Im1tt}), the flavored CP asymmetries $\epsilon^{\alpha\beta}_{1}$ are formed as,
\begin{equation}
\epsilon^{ee}_{1}=\frac{M_{1}(B^{L}_{1}B^{\phi}_{1})^{\frac{1}{2}}c_{ee}}{v^{2}[{\rm Tr}(m^{(1)\dagger}_{\nu}m^{(1)}_{\nu})]^{1/2}}
\Big[\frac{2}{3}z_{1}\left(z^{2}_{1}+z^{2}_{2}+2z_{1}z_{2}\cos\beta\right)^{\frac{1}{2}}
 \left(1-\varepsilon_{1}-\varepsilon_{2}\right)\sin(\beta-\sigma_{1})\Big],
 \label{eqn-CP1ee}
\end{equation}
\begin{eqnarray}
\nonumber
\epsilon^{e\mu}_{1}&\approx&\frac{M_{1}(B^{L}_{1}B^{\phi}_{1})^{\frac{1}{2}}c_{e\mu}}{v^{2}[{\rm Tr}(m^{(1)\dagger}_{\nu}m^{(1)}_{\nu})]^{1/2}}\Big[-
 \frac{2}{3}\varepsilon_{1}z_{1}\left(z^{2}_{1}+z^{2}_{2}+2z_{1}z_{2}\cos\beta\right)^{\frac{1}{2}}
 \sin(\beta-\sigma_{1})\Big]\\
 &\approx&
 \epsilon^{\mu e}_{1},
 \label{eqn-CP1em}
\end{eqnarray}
\begin{eqnarray}
 \nonumber
\epsilon^{\mu\mu}_{1}&\approx&\frac{M_{1}(B^{L}_{1}B^{\phi}_{1})^{\frac{1}{2}}c_{\mu\mu}}{v^{2}[{\rm Tr}(m^{(1)\dagger}_{\nu}m^{(1)}_{\nu})]^{1/2}}\times\varepsilon_{2}z_{1}\Big[\frac{1}{3}\left(z^{2}_{1}+z^{2}_{2}+2z_{1}z_{2}\cos\beta\right)^{\frac{1}{2}}\sin(\beta-\sigma_{1})\\
 &+&
 \left(z^{2}_{1}+z^{2}_{2}-2z_{1}z_{2}\cos\beta\right)^{\frac{1}{2}}\sin(\beta-\sigma_{2})\Big],
 \label{eqn-CP1mm}
\end{eqnarray}
and 
\begin{eqnarray}
\nonumber
\epsilon^{\tau\tau}_{1}&\approx&\frac{M_{1}(B^{L}_{1}B^{\phi}_{1})^{\frac{1}{2}}c_{\tau\tau}}{v^{2}[{\rm Tr}(m^{(1)\dagger}_{\nu}m^{(1)}_{\nu})]^{1/2}}\times \varepsilon_{1}z_{1}\Big[\frac{1}{3}\left(z^{2}_{1}+z^{2}_{2}+2z_{1}z_{2}\cos\beta\right)^{\frac{1}{2}}\sin(\beta-\sigma_{1})\\
 &+& \left(z^{2}_{1}+z^{2}_{2}-2z_{1}z_{2}\cos\beta\right)^{\frac{1}{2}}\sin(\beta-\sigma_{2})\Big].
 \label{eqn-CP1tt}
\end{eqnarray}

From equations (\ref{eqn-epsmat2}), (\ref{eqn-Im2ee}), (\ref{eqn-Im2em}), (\ref{eqn-Im2mm}) and (\ref{eqn-Im2tt}), the flavored CP asymmetries $\epsilon^{\alpha\beta}_{2}$ are formed as,
\begin{eqnarray}
\nonumber
\epsilon^{ee}_{2}&=&\frac{M_{2}(B^{L}_{2}B^{\phi}_{2})^{\frac{1}{2}}c_{ee}}{v^{2}[{\rm Tr}(m^{(2)\dagger}_{\nu}m^{(2)}_{\nu})]^{1/2}}\times-\frac{2}{3}z_{2}\times 
\\ \nonumber
&&\Big[\frac{2}{3}\left(z^{2}_{1}+z^{2}_{2}+2z_{1}z_{2}\cos\beta\right)^{\frac{1}{2}}(1-2\varepsilon_{1}-2\varepsilon_{2})\sin\sigma_{1}\\
 &+&\frac{1}{3}z_{1}(1+\varepsilon_{1}+\varepsilon_{2})\sin\beta\Big],
 \label{eqn-CP2ee}
\end{eqnarray}
\begin{eqnarray}
\nonumber
\epsilon^{e\mu}_{2}&\approx&\frac{M_{2}(B^{L}_{2}B^{\phi}_{2})^{\frac{1}{2}}c_{e\mu}}{v^{2}[{\rm Tr}(m^{(2)\dagger}_{\nu}m^{(2)}_{\nu})]^{1/2}}\times\frac{1}{3}z_{2}\times
\\
 \nonumber
 &&\Big[(\frac{7}{6}\varepsilon_{1}+\frac{1}{6}\varepsilon_{2}-\frac{1}{3})\left(z^{2}_{1}+z^{2}_{2}+2z_{1}z_{2}\cos\beta\right)^{\frac{1}{2}}\sin\sigma_{1}\\
 \nonumber
 &+&\frac{1}{3}(1+\varepsilon_{1}+\varepsilon_{2})z_{1}\sin\beta\\
 \nonumber
 &-&\frac{1}{2}(\varepsilon_{1}-\varepsilon_{2})\left(z^{2}_{1}+z^{2}_{2}-2z_{1}z_{2}\cos\beta\right)^{\frac{1}{2}}\sin\sigma_{2}\Big]\\
 &\approx&
 \epsilon^{\mu e}_{2},
 \label{eqn-CP2em}
\end{eqnarray}
\begin{eqnarray}
 \nonumber
\epsilon^{\mu\mu}_{2}&\approx&\frac{M_{2}(B^{L}_{2}B^{\phi}_{2})^{\frac{1}{2}}c_{\mu\mu}}{v^{2}[{\rm Tr}(m^{(2)\dagger}_{\nu}m^{(2)}_{\nu})]^{1/2}}\times-\frac{2}{3}z_{2}\times\\
 \nonumber 
 &&\Big[(-\frac{5}{6}\varepsilon_{1}+\frac{1}{6}\varepsilon_{2}-\frac{1}{6})\left(z^{2}_{1}+z^{2}_{2}+2z_{1}z_{2}\cos\beta\right)^{\frac{1}{2}}\sin\sigma_{1}\\
\nonumber
 &+&\frac{1}{3}(1+\varepsilon_{1}+\varepsilon_{2})z_{1}\sin\beta\\
 &+&\frac{1}{2}(1-\varepsilon_{1}-3\varepsilon_{2})\left(z^{2}_{1}+z^{2}_{2}-2z_{1}z_{2}\cos\beta\right)^{\frac{1}{2}}\sin\sigma_{2}\Big],
 \label{eqn-CP2mm}
\end{eqnarray}
and 
\begin{eqnarray}
\nonumber
\epsilon^{\tau\tau}_{2}&\approx&\frac{M_{2}(B^{L}_{2}B^{\phi}_{2})^{\frac{1}{2}}c_{\tau\tau}}{v^{2}[{\rm Tr}(m^{(2)\dagger}_{\nu}m^{(2)}_{\nu})]^{1/2}}\times-\frac{2}{3}z_{2}\times
\\
 \nonumber 
 &&\Big[(\frac{1}{6}\varepsilon_{1}-\frac{5}{6}\varepsilon_{2}+\frac{1}{6})\left(z^{2}_{1}+z^{2}_{2}+2z_{1}z_{2}\cos\beta\right)^{\frac{1}{2}}\sin\sigma_{1}\\
 \nonumber
 &+&\frac{1}{3}(1+\varepsilon_{1}+\varepsilon_{2})z_{1}\sin\beta\\
 &+&\frac{1}{2}(1-3\varepsilon_{1}-\varepsilon_{2})\left(z^{2}_{1}+z^{2}_{2}-2z_{1}z_{2}\cos\beta\right)^{\frac{1}{2}}\sin\sigma_{2}\Big].
 \label{eqn-CP2tt}
\end{eqnarray}

\begin{figure*}[htb!]
 \centering
\begin{subfigure}{0.45\linewidth}
 \includegraphics[width=\linewidth]{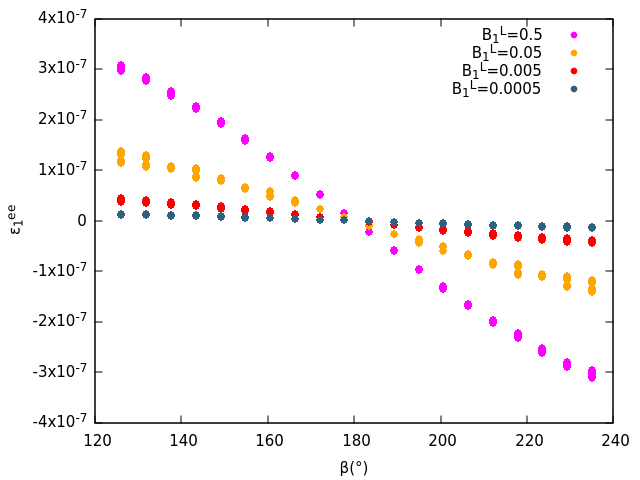}
\caption{$\epsilon^{ee}_{1}$ as a function of $\beta$}
\label{fig-BD11}
\end{subfigure}
 \begin{subfigure}{0.45\linewidth}
\includegraphics[width=\linewidth]{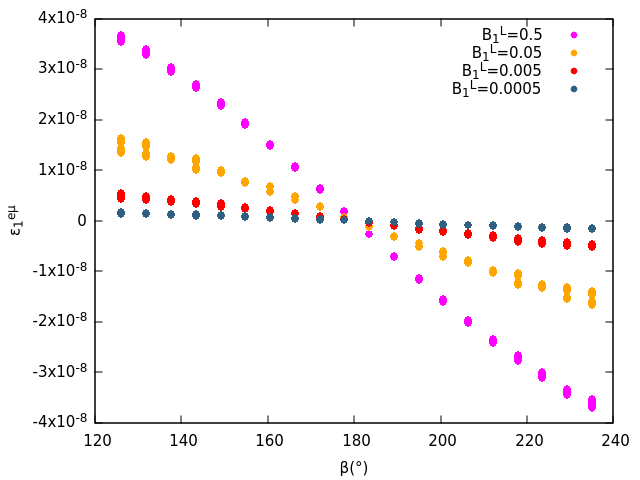}
\caption{$\epsilon^{e\mu}_{1}$ as a function of $\beta$}
\label{fig-BD12}
\end{subfigure}
\begin{subfigure}{0.45\linewidth}
\includegraphics[width=\linewidth]{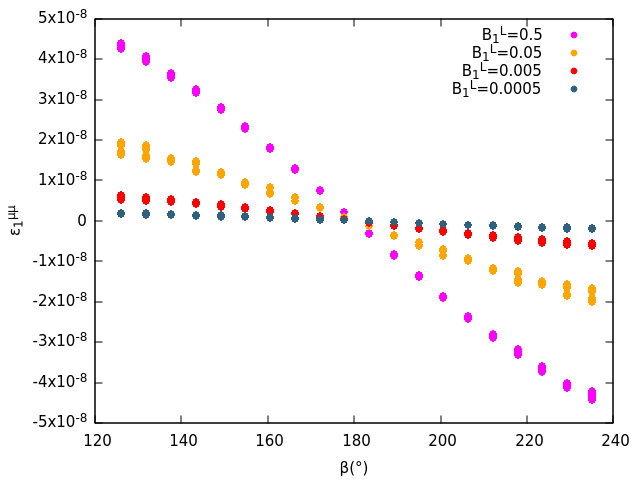}
\caption{$\epsilon^{\mu\mu}_{1}$ as a function of $\beta$}
\label{fig-BD13}
\end{subfigure}
\begin{subfigure}{0.45\linewidth}
\includegraphics[width=\linewidth]{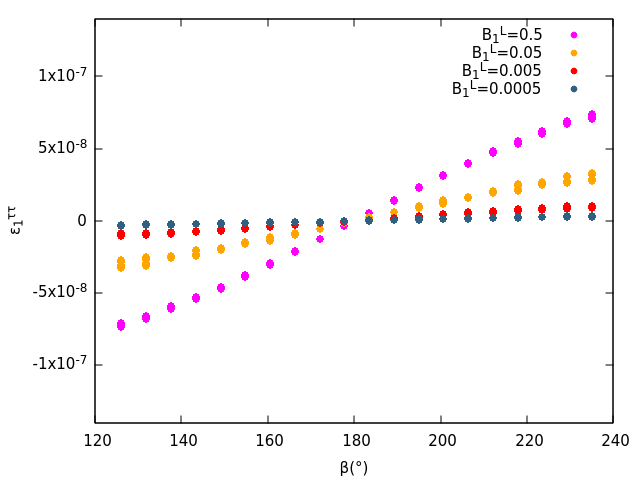}
\caption{$\epsilon^{\tau\tau}_{1}$ as a function of $\beta$}
\label{fig-BD14}
\end{subfigure}
\caption{Flavored CP asymmetries generated due to triplet $\Delta_{1}$ decays, as a function of phase $\beta$ for comparable and
hierarchical branching ratios. The triplet $\Delta_{1}$ mass $M_{1}$ is taken $2\times 10^{10}$ GeV.}
\label{fig-BD1}
\end{figure*}
\begin{figure*}[htb!]
 \centering
\begin{subfigure}{0.45\linewidth}
 \includegraphics[width=\linewidth]{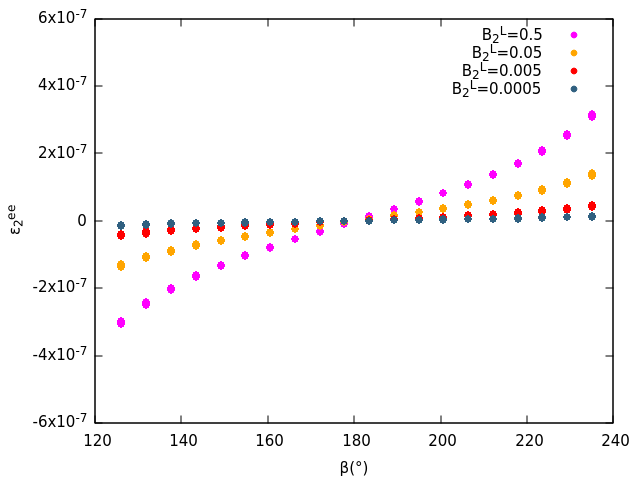}
\caption{$\epsilon^{ee}_{2}$ as a function of $\beta$}
\label{fig-BD21}
\end{subfigure}
 \begin{subfigure}{0.45\linewidth}
\includegraphics[width=\linewidth]{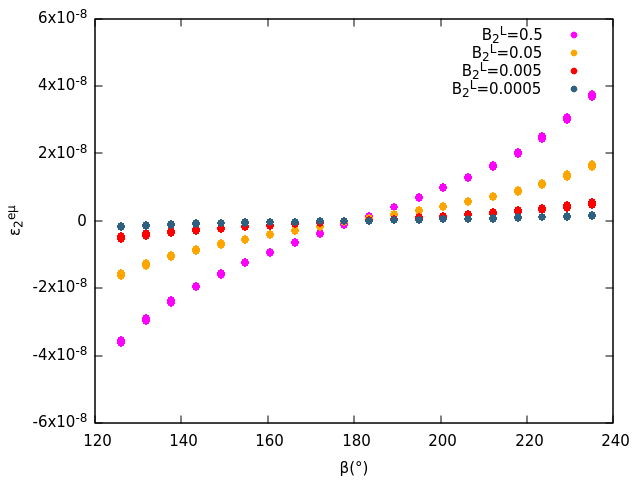}
\caption{$\epsilon^{e\mu}_{2}$ as a function of $\beta$}
\label{fig-BD22}
\end{subfigure}
\begin{subfigure}{0.45\linewidth}
\includegraphics[width=\linewidth]{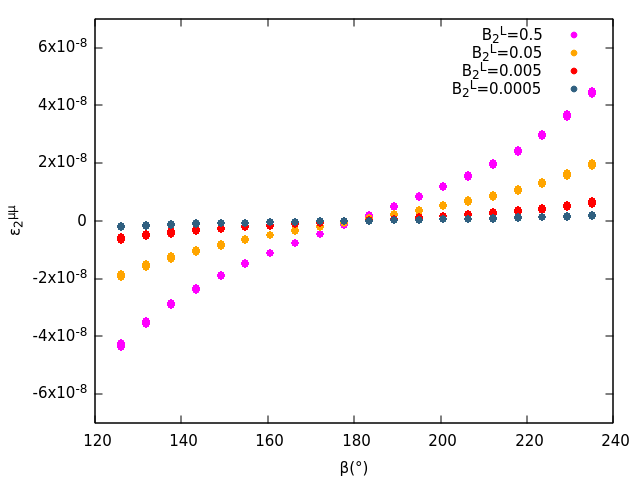}
\caption{$\epsilon^{\mu\mu}_{2}$ as a function of $\beta$}
\label{fig-BD23}
\end{subfigure}
\begin{subfigure}{0.45\linewidth}
\includegraphics[width=\linewidth]{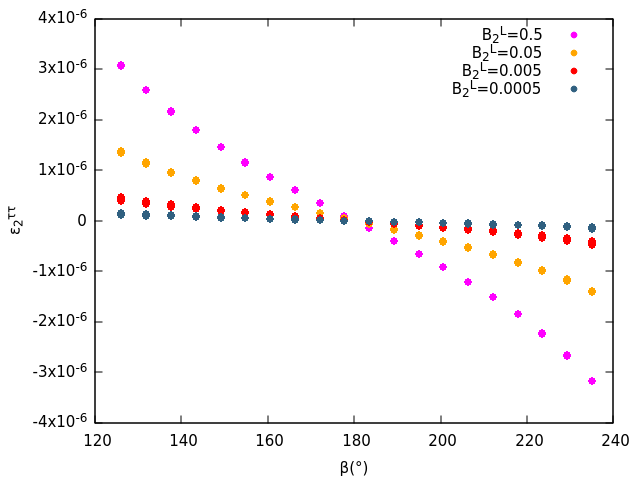}
\caption{$\epsilon^{\tau\tau}_{2}$ as a function of $\beta$}
\label{fig-BD24}
\end{subfigure}
\caption{Flavored CP asymmetries generated due to triplet $\Delta_{2}$ decays, as a function of phase $\beta$ for comparable and
hierarchical branching ratios. The triplet $\Delta_{2}$ mass $M_{2}$ is taken $2\times 10^{10}$ GeV.}
\label{fig-BD2}
\end{figure*}

\acknowledgments
SC acknowledges the valuable discussions with Rohan Pramanick on 
Spontaneous CP violation and leptogenesis.

\newpage
\bibliographystyle{plain}

\end{document}